\documentclass[preprint]{elsarticle}
\usepackage{amsthm}
\usepackage[cmex10]{amsmath}
\usepackage{amssymb}
\usepackage{dblfloatfix}
\usepackage{lineno,hyperref}
\usepackage[colorinlistoftodos]{todonotes}
\usepackage{stackrel}
\usepackage{pdfpages}
\usepackage{graphics}
\usepackage{epsfig}
\usepackage{times}
\usepackage{amsmath}
\usepackage{amssymb}
\usepackage{hyperref}
\usepackage{color}
\usepackage{arydshln}
\usepackage{arydshln}
\usepackage{gensymb}

\newcommand{\cF}{{\cal F}}
\newcommand{\bR}{{\bf R}}

\newcommand{\cA}{{\cal A}}

\newcommand{\cT}{{\cal T}}

\newcommand{\mF}{{\bf F}}
\newcommand{\mZ}{{\bf Z}}
\newcommand{\bx}{{\bf x}}
\newcommand{\by}{{\bf y}}
\newcommand{\bz}{{\bf z}}
\newcommand{\ba}{{\bf a}}
\newcommand{\beff}{{\bf f}}

\newcommand{\cC}{{\cal C}}
\newcommand{\cS}{{\cal S}}

\newcommand{\bitem}{\begin{itemize}}
\newcommand{\eitem}{\end{itemize}}
\newcommand{\goto}{\rightarrow}
\newcommand{\beq}{\begin{equation}}
\newcommand{\eeq}{\end{equation}}

\newcommand{\1}{ {\bf 1} }
\newcommand{\bi}{ {\bf i} }

\newcommand{\mR}{\mathbb{R}}

\newcommand{\mH}{\mathbb{H}}

\newcommand\card{\text{card}}

\newtheorem{thm}{Theorem}[section]

\newtheorem{lem}[thm]{Lemma}

\newtheorem{dfn}[section]{Definition}

\newcounter{comment}

\nolinenumbers

\journal{arXiv}

\begin{document}

\begin{frontmatter}

\title{Incoherence of Partial-Component Sampling in multidimensional NMR}

\author[mymainaddress]{Hatef Monajemi\corref{mycorrespondingauthor}}
\cortext[mycorrespondingauthor]{Corresponding author}
\ead{monajemi@stanford.edu}

\author[mymainaddress]{David L. Donoho}
\author[mysecondaddress]{Jeffrey C. Hoch}
\author[mysecondaddress]{Adam D. Schuyler}

\address[mymainaddress]{Department of Statistics, Stanford University, CA, 94305}
\address[mysecondaddress]{Department of Molecular Biology and Biophysics, University of Connecticut Health Center, CT, 06030}

\begin{abstract}
In NMR spectroscopy, undersampling in the indirect dimensions causes reconstruction artifacts whose size can be bounded
using the so-called {\it coherence}. In experiments with multiple indirect dimensions, new undersampling approaches were recently proposed: random phase detection (RPD) \cite{Maciejewski11} and its generalization, partial component sampling (PCS) \cite{Schuyler13}. The new approaches are fully aware of the fact that high-dimensional experiments generate hypercomplex-valued free induction decays; they randomly acquire only certain low-dimensional components of each high-dimensional hypercomplex entry. We provide a classification of various hypercomplex-aware undersampling schemes, and define a hypercomplex-aware coherence appropriate for such undersampling schemes; we then use it to quantify undersampling artifacts of RPD and various PCS schemes.
\end{abstract}


\end{frontmatter}



\section{Introduction}

In traditional NMR spectroscopy, the complete dataset covers
a grid $(t_1,\dots ,t_d)$, where $t_d$ varies along the direct (\textit{a.k.a} acquisition) time dimension and $(t_1,\dots, t_{d-1})$ along the indirect time dimensions, which are sampled parametrically by separate experiments. Following \cite{Barna86},
many researchers achieved acceptable reconstruction with non-uniform sampling (NUS), in which 
they acquired only a scattered subset of the indirect times. Typically the
 undersampling scheme involved either random uniform sampling
 or Poisson sampling with an exponentially decaying rate
function\cite{hoch13}. In many cases an NUS experiment can save a great deal of experiment time, while
still producing an acceptable result.

Reconstructions from undersampled data will in general display artifacts,
and it is important to understand and quantify them in order to know if the
reconstructions are acceptable despite  undersampling.
{\it Coherence} provides a useful bound on the size of undersampling artifacts;
it has been applied in NMR spectroscopy and MR imaging \cite{hoch13, CSMRI}, and in the mathematical study of compressed sensing \cite{DonohoHuo, DonohoElad02, Tropp04}. Conceptually, coherence bounds the extent to which a point mass in the true
underlying spectrum at any one $k$-tuple can generate apparent mass in the reconstruction at some other $k' \neq k$.
By controlling coherence we keep artifacts small. In fields outside NMR spectroscopy,
signals are either real or complex valued, and acceptable definitions of coherence
have been proposed and applied. However, these definitions are not specifically adopted for the NMR spectroscopy.

When States-Haberkorn-Ruben phase-sensitive detection (PSD) \cite{States82} 
is employed for frequency sign discrimination in the indirect dimensions, 
the complete data at each $t$-tuple $(t_1,\dots ,t_d)$
in a $d$-dimensional experiment are 
$2^{d}$-dimensional hypercomplex numbers produced by $2^{d-1}$ complex reads. The required algebra of such hypercomplex numbers 
has been described in detail by Delsuc \cite{Delsuc88}. 
NUS can be applied in a fashion that is ignorant of
the hypercomplex structure, and the traditional definition 
of coherence can be straightforwardly generalized for NUS in the hypercomplex case.

Recently, novel hypercomplex-aware undersampling schemes were proposed for the multidimensional case;
for example, in a $d$-dimensional experiment, RPD \cite{Maciejewski11} acquires at a given $t$-tuple only a single complex measurement
rather than all $2^{d-1}$ complex reads. More generally, PCS \cite{Schuyler13}
acquires at a given $t$-tuple, $2m$ components of a hypercomplex datum produced by $1\leq m \leq 2^{d-1}$ complex reads.  
In the most general PCS scheme, the number of components being sampled may even jump around somewhat randomly between $t$-tuples. To understand the artifacts caused by such undersampling,
this paper formulates a hypercomplex-aware definition of coherence.
This new definition, although somewhat more involved than the traditional
non-hypercomplex-aware quantity, seems to be the right notion for
studying RPD and PCS because it obeys the core exact-reconstruction result that one wants
coherence to obey: As we show here, a sufficiently low coherence  
allows a sufficiently sparse spectrum to be recovered correctly
by convex optimization.

\section{Approaches to Undersampling in NMR}

We call each specific tuple $(t_1,t_2,\dots,t_{d-1})$ of indirect sampling times an \textit{indel} (for {\it ind}irect {\it el}ement).
Associated with each indel is a hypercomplex-valued free induction decay (FID) 
$f_{t_1,\dots,t_{d-1}} = \{f_{t_1,\dots,t_{d-1}}(t_d): 0 \leq t_d < T_d\}$. 
Each hypercomplex entry $ f_{t_1,\dots,t_{d-1}}(t_d)$ can be represented \-- see farther below \-- 
as a tuple of $m=2^{d-1}$ complex numbers $f = (f^1,\dots,f^{m})$ and so one can 
equivalently view the indel as generating a set of $m$ complex-valued FID's $f^j_{t_1,\dots,t_{d-1}} = \{ f^j_{t_1,\dots,t_{d-1}}(t_d): 0 \leq t_d < T_d \}$ for $j=1,\dots,m$.

In many well-known  applications of undersampling in multidimensional NMR \cite{Bodenhausen81,Barna86,hoch-book}, 
the indels have been sampled nonuniformly, however, 
for each sampled indel, $(t_1,t_2,\dots,t_{d-1})$  say, the {\it full} hypercomplex FID 
$f_{t_1,\dots,t_{d-1}}$ is acquired. 

\newcommand{\cJ}{{\cal J}}
In the recent RPD proposal,
undersampling is effected  by partial sampling of the hypercomplex FID. 
One samples indels {\it exhaustively},
but has a {\it single-component sampling schedule} $\cJ = \{ j(t_1,t_2,\dots,t_{d-1}) : 0 \leq t_k  < T_k, k=1,\dots, d \}$ 
specifying which single complex component of the FID to sample
at each specific indel. Namely, at indel
$(t_1,t_2,\dots,t_{d-1})$, one acquires only the complex-valued 
FID $f^j_{t_1,\dots,t_{d-1}}$. This allows for undersampling by a 
factor $m = 2^{d-1}$ in a $d$-dimensional experiment.
In the simplest variant of the original proposal, the sampling schedule
selects the sampled coordinate at random indel-by-indel.

In the more general PCS proposal 
(see Schuyler et al. \cite{Schuyler13,Schuyler15}), 
one specifies a {\it component-subset sampling schedule}. 
$\cJ = \{ J(t_1,t_2,\dots,t_{d-1}) : 0 \leq t_k  < T_k, k=1,\dots, d \}$.
Here each $J$ specifies the indices corresponding to a subset of the $m$ coordinates $(x_1,\dots,x_m)$
of the full hypercomplex FID. Hence if a specific indel has $J(t_1,t_2,\dots,t_{d-1}) = \{1,3\}$, the
experiment will acquire the two complex FID's $f^1_{t_1,t_2,\dots,t_{d-1}}$ and $f^3_{t_1,t_2,\dots,t_{d-1}}$.
In the special case where the selected subset is always a singleton at each indel, we recover RPD.
In the simplest case, PCS uses a subset of the same \emph{cardinality}, say $S$, at each indel, however
varying the subset from one indel to the next,
in a random fashion. This allows for undersampling by a 
factor $2^{d-1}/S$ in a $d$-dimensional experiment,
for each $S=1,2,\dots,2^{d-1}$. 

One may combine partial sampling of indels with partial sampling of hypercomplex components.
Notationally, one simply extends the notion of subset-sampling schedule
to allow empty sets at certain indels, and nonempty sets at others.
It is easy to
visualize PCS schemes with fixed cardinality of component subset for sampled indels; for example,
a 3-dimensional experiment measuring 2 complex FID's at
all indels, rather than 4, thereby saving 50\% on the measuring time. If one further samples only half of the indels, the saving increases to 75\%. 
Let's let FCPCS stand for such {\it fixed-cardinality  partial component sampling} schemes. 

\begin{figure*}[h]
\centering
$
\includegraphics[width=6in]{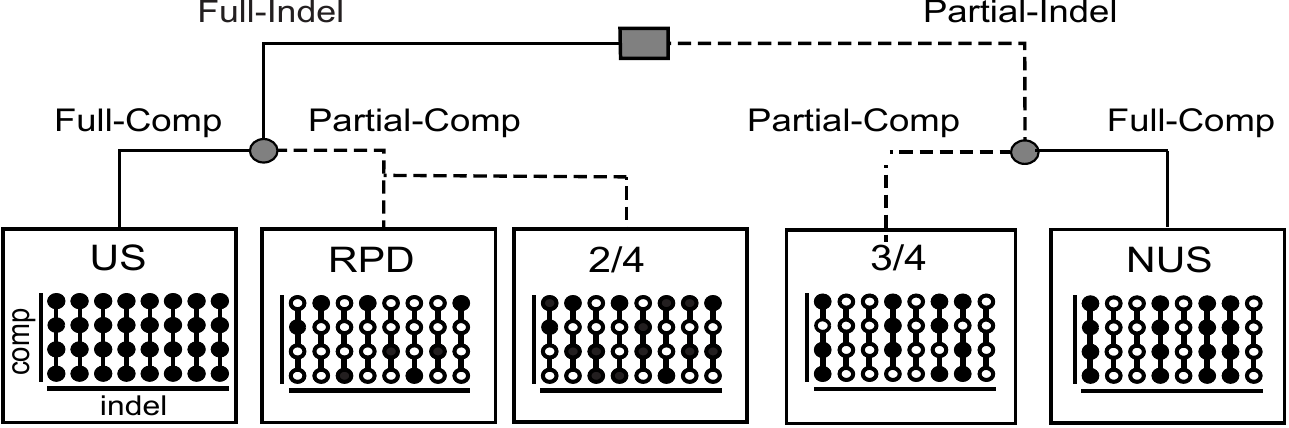}
$
\caption{Examples of sampling schedules in multi-dimensional NMR. The top level branch indicates full (solid lines) versus partial (dashed lines) sampling of indels. Each of the resulting two branches reaches a fork (gray circles) at which we specify which components are being sampled at each indel: full-component (solid lines) versus fixed-cardinality partial-component (dashed lines) sampling.  Each panel illustrates a collection of indels each represented as a vertical series of connected circles, with each circle corresponding to a complex read. Black (white) filled circles indicate components that are collected (omitted).}
\label{sampling-tree}
\end{figure*}

Figure \ref{sampling-tree} may help the reader to envision some
of the possibilities envisioned, and how they accommodate existing approaches
that the reader would already be familiar with. 

The five possibilities indicated in Figure 1 are
\begin{itemize}
\item[+] {\bf Uniform sampling (US)} in which indels and their hypercomplex components are both fully sampled. 
\item[+] {\bf Nonuniform sampling (NUS)} in which all hypercomplex components are sampled for nonuniformly sampled indels.
\item[+] {\bf Random phase detection (RPD)} in which only one complex FID is sampled at each indel.
\item[+] {\bf Full Indel/2-Component sampling} in which 
all indels are sampled, but only two complex FID's are sampled 
at those indels.
\item[+] {\bf Partial Indel/3-Component sampling} in which 
only some indels are sampled, at which we acquire three
complex FID's.


\end{itemize}

\begin{figure*}[h]
\centering
$
\includegraphics[width=2in]{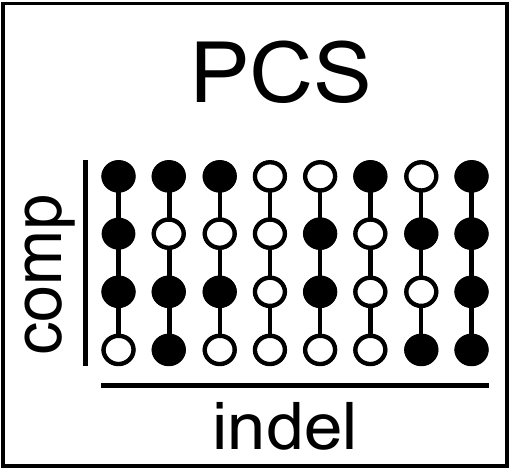}
$
\caption{PCS in its most general form can acquire subsets of components of arbitrary dimension varying indel-by-indel. This includes
all the simpler schemes shown in Figure \ref{sampling-tree}.}
\label{general-pcs}
\end{figure*}

At an abstract level all five possibilities indicated in this figure are simply special cases of 
the fully-general notion of PCS,
however for readers making their first acquaintance with our notation, it seems helpful
to keep these special cases at the front of the reader's mind.
Figure \ref{general-pcs} shows a fully general Partial-Component sampling approach
in which we make measurements of varying cardinality, indel by indel.

\section{The traditional measure of incoherence}
The {\it point-spread function} (PSF) and the {\it peak-to-sidelobe} ratio (PSR) are traditional
signal processing concepts that quantify the
extent to which a true underlying `spike' might, upon reconstruction,
appear to `leak' to other locations. They were used in MR imaging and 
spectroscopy to assess undersampling artifacts \cite{hoch13,CSMRI}. 

In traditional   
signal processing\footnote{Examples range from radar to  acoustic and wireless signal processing.}
there is an underlying vector $\bx$ 
of interest, we acquire  
a vector $\by$ of complex-valued measurements 
according to the matrix equation $\by = A \bx$, 
with $A$ a matrix having complex-valued entries.
Each column $\ba_i$ of $A$ represents the data acquired by a unit 
spike located in vector $\bx$ at coordinate $i$.  The classical matched-filter
reconstruction is 
\[
   \hat{x} = D^{-1} A^* \by,
\]
where $A^*$ is the Hermitian transpose of $A$ and $D = diag(A^*A)$
is a normalizing operator. In the special case
where we were trying to recover a point mass signal located at $i_0$,
then $\bx_i = \delta_{i_0}(i)$, and
the formula gives $\hat{x}_{i_0} = 1$ as we might hope.
however, we would not be so lucky as to also
have $\hat{x}_j = 0$, $j \neq i_0$; 
the point mass would be spread out. 
Define the (normalized) {\it point spread function }
\[
PSF(i,j) = \frac{(A^*A)_{i,j}}{(A^*A)_{i,i}},
\]
For a general spike located at position $i$, $x^{i}_j = \delta_{i}(j)$, and generating data $\by^i = A \bx^i$
and matched-filter reconstruction $\hat{\bx}^{i} = D^{-1} A^* \by^i$, we have that 
$PSF(i,j) = \hat{x}^{i}_j$.
We therefore quantify the ability of matched filtering
to sharply recover a point mass by the size of the maximum sidelobe:
 $MS = \max_{j \neq i}|PSF(i,j)|$. For example if this quantity were zero,
 then necessarily we would have a perfect reconstruction: $\hat{\bx}^i \equiv \bx^i$. 
 Some authors consider
 peak-to-sidelobe ratio
\[
   PSR = \min_{j\neq i} \frac{PSF(i,i)}{PSF(i,j)},
\]
which is reciprocal to the sidelobe height (as the peak is normalized to unit height).

In the literature on undersampling and compressed sensing,
an equivalent notion is called coherence.
We assume that the matrix $A$ has columns $\ba_j$ of unit $\ell_2$ norm,
and then define 
%
the coherence $\mu$\cite{DonohoHuo}   
\begin{equation} \label{eq:cohdef}
\mu = \max_{i,j, i\neq j} | \ba_i^* \ba_j |.
\end{equation}
When $A$ is normalized in this way, $|PSF(i,i)|=1$,
and so $\mu = MS = (PSR)^{-1}$.

These notions apply immediately to the undersampled multidimensional NMR situation,
at the cost of some explanation. 
We can represent the complete noiseless spectrum
as a vector $\bx = (x_k)$, where $k$ runs through all the $k$-space indices underlying the spectrum
and all the complex components of the hypercomplex-valued spectrum.
We can represent the collection of all complex-valued
samples obtained in an experiment as a vector $\by = (y_i)$ where the index $i$
runs through an enumeration of all the indels and hypercomplex components that were sampled.
The mapping $\bx \mapsto \by$ is complex linear, and thus can be 
represented by a complex matrix $A$.
To describe the matrix $A$ more concretely we develop more terminology and machinery.

\section{Hypercomplex algebra for multidimensional NMR}
As mentioned above, the signal acquired in a multi-dimensional 
NMR experiment is hypercomplex \cite{hoch-book,States82,Delsuc88,Schuyler13}.  
The hypercomplex algebra $\mH_d$ used in multi-dimensional 
NMR is the algebra defined over the real field $\mR$ 
with $d$ generators satisfying the following relations:
\bitem
\item $\bi_j^2 = -1,  \quad j = 1,2,\dots, d$,
\item $\bi_j \cdot \bi_{k} = \bi_{k} \cdot \bi_{j}$. 
\eitem
With this definition, the algebra is \emph{commutative} \cite{Delsuc88, Schuyler13}. 
These generators produce $2^d$ basis elements:
$$\lbrace {\bf 1}, \bi_1,...,\bi_d , \bi_{1,2},  \bi_{1,3}, \dots, \bi_{d-1,d} , \dots , \bi_{1,2, \dots ,d} \rbrace.$$
Each subset of $\{\bi_1,\dots,\bi_d\}$ corresponds to one
basis element;  the rule for obtaining $\bi_{j_1,j_2,\dots,j_k}$ is: `turn commas
into multiplications': $\bi_{j_1,j_2,\dots,j_k} \equiv \bi_{j_1} \cdot \bi_{j_2} \dots \bi_{j_k}$.
This basis can 
represent any element of the $2^d$-dimensional vector space $\mH_d$ as a linear combination
\begin{equation*}
z  = z_0 +  z_1 \bi_1 +  \dots + z_{d+1} \bi_{1,2}  + \dots + z_{2^d-1} \bi_{1,2, \dots ,d}.
\end{equation*}
Here the coefficients $z_j \in \mR$ for $j = 0,\dots, 2^d-1$. 

\subsection{Basic notions}

Define the mapping $\phi: \mH_d \goto \mR^{2^d}$ that takes a hypercomplex element
and delivers its underlying $2^d$-vector of coefficients. For instance, 
$$
\phi(z_0 +  z_1 \bi_1 +  \dots + z_{d+1} \bi_{1,2}  + \dots + z_{2^d-1} \bi_{1,2, \dots d}) = \left( z_0 ,  z_1 , \dots , z_{d+1} ,\dots , z_{2^d-1}\right)^T.
$$ 

\begin{itemize}

\item {\it Real and Complex Elements.} We call an element $z$ {\it real} if $\phi(z) = (z_0, 0,0,0,\dots,0)$.
Let $\bi \in \{\bi_1, \dots \bi_d \}$ be a generator 
and call any element $z = a + b \bi$, where $a$ and $b$ are real,
a `complex' element. Evidently for a complex $z = a + b \bi_j$, $\phi(z)_0 = a$ and $\phi(z)_j = b$,
while all other entries of $\phi(z)$ vanish.
Note that here and below, we abuse terminology and make no distinction between the traditional real $a \in \bR$ and the real
element $a \cdot {\bf 1} \in \mH_d$ with 
$\phi(a \cdot {\bf 1})  = (a,0,0,\dots,0)$, though of course the two objects live in different spaces (i.e. $\bR$ vs. $\mH_d$); the reader is expected to work out our intent from context.

\item {\itshape Hypercomplex modulus}: 
We define the modulus of a hypercomplex number 
as the usual two-norm of the corresponding real vector, namely 
$$
| z |_\mH  =  \|\phi(z)\|_2=   \sqrt{\sum_{j = 0}^{2^d-1} z_j^2} .
$$ 
In this way we make $\mH_d$ isometric to $\mR^{2^d}$.
Also we call $\phi$ the {\it vector isomorphism}, i.e.
the isomorphism between $\mH_d$ and the vector space $\mR^{2^d}$.

\item{\itshape Hypercomplex conjugation:}
The conjugate of $z \in \mH_d$ is given by, 
\begin{eqnarray*}
z^\sharp &=& z_0 + (-1)^{\kappa(\bi_1)} z_1 \bi_1 +  \dots \\
&&+ (-1)^{\kappa(\bi_{1,2})} z_{d+1} \bi_{1,2}  + \dots \\
&&+ (-1)^{\kappa(\bi_{1,2, \dots, d})} z_{2^d-1} \bi_{1,2, \dots,d}.
\end{eqnarray*}
where 
$\kappa(\bi_{g}) =  \{\card(g) \bmod 2\}$ for index group $g$. Namely, $\kappa = 0$ if the number of elements of the index group is even and   $\kappa = 1$ otherwise. For instance, an element of $\mH_3$ takes the following representation:
\begin{eqnarray*}
z  &=& z_0 +  z_1 \bi_1 + z_2 \bi_2 + z_3 \bi_3 \\
&&+z_{4} \bi_{1,2} + z_{5} \bi_{1,3}+ z_{6} \bi_{2,3} +z_{7} \bi_{1,2,3}\\
z^\sharp  &=& z_0 -  z_1 \bi_1 - z_2 \bi_2 - z_3 \bi_3 \\
&&+ z_{4} \bi_{1,2} + z_{5} \bi_{1,3}+ z_{6} \bi_{2,3} -z_{7} \bi_{1,2,3}.
\end{eqnarray*}
Since $z \mapsto z^\sharp$ preserves the
absolute value of each real coefficient, we have
$\| \phi(z)\|_2 = \|\phi(z^\sharp)\|_2$ and 
so $|z|_{\mH} = |z^\sharp|_{\mH}$; conjugation is an isometry.
Notice in particular that for a generator $\bi \in \{\bi_1,\dots, \bi_d \}$,
$\bi^\sharp = - \bi$. Since for each generator $\bi^2 = -1$, it follows that
$\bi \cdot \bi^\sharp=1$.


\item {\itshape Factorizable hypercomplex number}: 

We say that the hypercomplex element $x \in \mH_d$ is
\emph{factorizable} if it can be represented as product of $d$ complex elements \-- i.e., if it may be written\footnote{In particular, not all elements can be written in this way;
Factorizable elements obey $z_g= (\prod_{j \in g} b_j)(\prod_{j \not \in g} a_j)$
for appropriate coefficients $(a_j,b_j)$.},\footnote{In the case of $d=2$, Delsuc  \cite{Delsuc88} calls such elements ``bi-complex''.}  
$$
x = (a_1+b_1 \bi_1) \cdot (a_2+b_2 \bi_2) \cdots  (a_d+b_d \bi_d).
$$
Let $\mF_d \subset \mH_d$ denote the set of factorizable elements.
For factorizable $x \in \mF_d$, one can verify that $x^\sharp \cdot x\in \mH_d$ is actually `real',
namely $\phi(x^\sharp \cdot x)$ has only its first entry nonzero.
In fact $\phi(x^\sharp \cdot x) = (|x|_\mH^2,0,0,0,\dots,0)$, and the norm factors as well:
$$
 |x|_\mH^2 = \prod_{j=1}^d (a_j^2+b_j^2).
$$
For factorizable elements, we can therefore abuse notation by writing $x^\sharp \cdot x = |x|_\mH^2$.
However, this identity is in general not true 
for an arbitrary $z \in \mH_d$. For a factorizable hypercomplex $x \in \mF_d$, one can define a multiplicative inverse:
$$
x^{-1} = \frac{x^\sharp}{x^\sharp \cdot x}  = \frac{x^\sharp}{|x|_H^2}. 
$$
 \end{itemize}
 Again, for non-factorizable elements $z \in \mH_d \backslash \mF_d$ no such identity holds in general.

\subsection{Matrix isomorphism}
The hypercomplex algebra $\mH_d$ is isomorphic to a subalgebra of the
algebra\footnote{i.e. the so-called total matrix algebra} of $2^d$-by-$2^d$ matrices with real entries $M_{2^d,2^d}$ \cite{Abian71}. 
We define a mapping $\Phi: \mH_d \mapsto M_{2^d,2^d}$ implementing this isomorphism
first of all, by its action on generators. Define two special 2-by-2 matrices
\begin{eqnarray*}
\1_c = \left(
\begin{array}{cc}
1 & 0  \\
0 & 1   
\end{array} 
\right)
& \text{and} & 
\bi_c = \left(
\begin{array}{cc}
0 & -1  \\
1 & 0   
\end{array} \right).
\end{eqnarray*}
and let $\otimes$ denote the Kronecker (tensor) product of matrices,
such that $\1_c \otimes \1_c$ is a 4-by-4 matrix (in fact, the 4-by-4 identity),
while \[ \underbrace{\1_c \otimes \1_c \otimes \dots \otimes \1_c}_d\] is
a $2^d$-by-$2^d$ matrix (this-time the $2^d$-by-$2^d$ identity, ${\bf 1}$, say). 
With this machinery in place, define for each generator $\bi_j$,  $ j= 1,2,\dots,d$,
\begin{equation*}
\Phi(\bi_j ) = \underbrace{\underbrace{\1_c \otimes \1_c \otimes \dots \1_c}_{d-j} \otimes \ \bi_c \otimes \underbrace{\1_c \otimes \dots \1_c}_{j-1} }_d.
\end{equation*}
Also define $\Phi(1)={\bf 1}$, where the argument $1$ denotes the real element of $\mH_d$ with $\phi(1) = (1,0,\dots,0)$
and the value of $\Phi(1)$ is the identity matrix ${\bf 1}$.

The other basis elements are induced from these $d$ isomorphic matrix generators of the algebra
using the principle of homomorphism. For example we define
\[
  \Phi(\bi_{j,k}) = \Phi(\bi_{j} \cdot \bi_{k}) = \Phi(\bi_{j} ) \cdot \Phi( \bi_{k}),
\]
The product $\bi_{j} \cdot \bi_{k}$ {\it inside } $\Phi$ is taking place in $\mH_d$,
while the product $\Phi(\bi_{j} ) \cdot \Phi( \bi_{k})$ {\it outside} $\Phi$ is taking place in $M_{2^d,2^d}$.
We then define the $2^d\times 2^d$ matrix isomorphism corresponding to an arbitrary element of the algebra $z \in \mH_d$ as, simply,
\begin{eqnarray}
\Phi(z)  &=& z_0 \Phi(\1)  +  z_1 \Phi(\bi_1) +  \dots + z_{d+1} \Phi(\bi_{1,2})  \nonumber
\\&&
+ \dots + z_{2^d-1} \Phi(\bi_{1,2,\dots, d}). \label{def:matrixiso}
\end{eqnarray}
As an example, the identity and 3 generators of  $\mH_3$ are given by
\begin{align*}
\Phi(\1) =  \1_c  \otimes \1_c \otimes \1_c, \quad \Phi(\bi_1) = \1_c \otimes \1_c \otimes \bi_c \\
\Phi(\bi_2) = \1_c \otimes \bi_c \otimes \1_c, \quad
\Phi(\bi_3) =  \bi_c \otimes \1_c \otimes \1_c
\end{align*}
The other elements of $\mH_3$ are produced accordingly. For instance,

\begin{align*}
\Phi(\bi_{1,2}) = \Phi(\bi_1 \cdot \bi_2) = \Phi(\bi_1) \Phi(\bi_2) =  \1_c  \otimes \bi_c \otimes \bi_c, \\
\Phi(\bi_{1,2,3}) =  \Phi(\bi_{1,2} \cdot \bi_3) = \Phi(\bi_{1,2}) \Phi(\bi_3) = \bi_c \otimes \bi_c \otimes \bi_c
\end{align*}
The reader might have already noticed the pattern inherent in producing the basis elements. Namely, $\1_c$'s are replaced by $\bi_c$ at locations dictated by the index group of the basis elements. 

One verifies that $\Phi$ respects conjugation in the two algebras,  $\Phi(z^\sharp) = \Phi(z)^T$,
for all $z \in \mH_d$, by checking that $\Phi(\bi^\sharp) = \Phi(\bi)^T$ on generators $\bi$.

\subsection{Hypercomplex multiplication as matrix-vector product}
For $x,y \in \mH_d$, one can verify that 
$$
\phi(x \cdot y)  = \Phi(x) \phi(y) = \Phi(y) \phi(x).
$$
Here, $\Phi : \mH_d \goto M_{2^d,2^d}$ denotes the previously defined matrix isomorphism, while $\phi:\mH_d \goto \mR^{2^d}$
denotes the vector isomorphism.  
Note that there is a one-to-one mapping between $\Phi(x)$ and $\phi(x)$, because
of (\ref{def:matrixiso}). For instance, check the following for $\mH_2$:
$$
\Phi(x) = \left[\phi(x),   \Phi(\bi_1)  \phi(x), \Phi(\bi_2) \phi(x), \Phi(\bi_{1,2}) \phi(x)\right].
$$
\subsection{Hypercomplex Fourier Transform}

Let $z = (z_{t_1,\dots,t_d}: 0 \leq t_j < T_j, j=1,\dots, d)$ denote a $d$-dimensional hypercomplex array,
with extent $T_j$ in dimension $j$ for $j=1,\dots, d$,
and let $\cA(\mH_d,T_1,\dots,T_d)$ denote the collection of all such arrays.
We are now in a position to define the hypercomplex Fourier transform $\cF_d$ as a linear mapping
from $\cA(\mH_d,T_1,\dots,T_d)$ to $\cA(\mH_d,T_1,\dots,T_d)$ \cite{Delsuc88,Schuyler13}.

For clarity in the next few paragraphs, let $\exp_{\mH}$ denote
the exponential in $\mH_d$. Let $\theta$ be real and $\bi$ be a generator.
We can  understand the expression  $\exp_{\mH}( \theta \bi)$
abstractly as a power series in $\mH_d$ 
\begin{eqnarray}
\exp_{\mH}(\theta  \bi ) &=& \sum_{\ell=0}^\infty \theta^\ell \bi^\ell / {{\ell}!} . 
\end{eqnarray}
For generator $\bi$, we have  $\bi^{2k} = (-1)^{k} {\bf 1}$, for integer $k \geq 1$,
so 
\begin{equation} \label{eq:exp-generator}
\exp_{\mH}(\theta \bi) = \cos_{\mR}(\theta){\bf 1} + \sin_{\mR}(\theta) \bi,
\end{equation}
where on the left, $\exp_{\mH}$ denotes the hypercomplex exponential,
while on the right, $\cos_{\mR}$ and $\sin_{\mR}$ denote the classical 
real-valued trigonmetric functions.
In particular $\exp_{\mH}(\theta \bi)$ is a complex element of $\mH_d$ (and is unimodular). From
$ \exp_{\mH}(\theta \bi)^\sharp = \exp_{\mH}(\theta \bi^\sharp) = \exp_{\mH}(-\theta \bi)$, and from the
fact that $\exp_{\mH}(-\theta \bi) = \cos_{\mR}(\theta) {\bf 1}- \sin_{\mR}(\theta) \bi$, we obtain 
$\exp_{\mH}(\theta \bi ) \exp_{\mH}(-\theta \bi )=1$.


For $k \in \mZ$  and $\bi$ a generator, (\ref{eq:exp-generator}) specializes to
$\exp_{\mH}( \frac{2 \pi t}{T} k \bi ) = \cos_{\mR}(\frac{2 \pi t}{T} k){\bf 1} + \sin_{\mR}(\frac{2 \pi t}{T} k) \bi$.
From the classical exponential sum over $\mR$ we get the following exponential sum over $\mH_d$:
\begin{equation} \label{eq:exponsum}
     \sum_{t=0}^{T-1} \exp_{\mH}( \frac{2 \pi t}{T} k \bi )  = \left\{ \begin{array}{ll} T  & k =0  \\
                 0 & k \ne 0\end{array} \right . .
\end{equation}
(This equation demands special care the first time one sees it.
Since the left side belongs to $\mH_d$, 
the right side denotes a so-called {\it real element}  $z \in \mH_d$, of the form $z = (z_0,0,\dots,0)$, with $z_0=T$ or $z_0 = 0$.)
This extends immediately to a multivariate exponential sum over $\mH_d$
\begin{equation} \label{eq:multisum}
     \sum_{t_1=0}^{T_1-1} \cdots \sum_{t_d=0}^{T_d-1} \prod_j \exp_{\mH_d}( \frac{2 \pi t_j}{T_j} k_j \bi_j )  = \left\{ \begin{array}{ll} \prod_j T_j  & k_1=\dots k_d =0  \\
                 0 & k \ne 0\end{array} \right . .
\end{equation}
(Again the right side is a real element of $\mH_d$).

We can also understand the exponential over $\mH_d$
using the Matrix isomorphism $\Phi$ and the exponential $\exp_{M_{2^d,2^d}}$ of matrices $M_{2^d,2^d}$; 
\[ 
\Phi(\exp_{\mH_d}(z)) = \exp_{M_{2^d,2^d}}(\Phi(z)).
\] 
From the commutativity  $\Phi(\bi_i)\Phi(\bi_j) = \Phi(\bi_j)\Phi(\bi_i)$
with   $\bi_i$ and $\bi_j$  generators,
we obtain for the matrix exponential 
\[
\exp_{M_{2^d,2^d}}(\Phi(\theta_1 \bi_1 + \theta_2 \bi_2 + \dots + \theta_d \bi_d) ) = \prod_j \exp_{M_{2^d,2^d}}(\Phi(\theta_j \bi_j)),
\]
and hence for the exponential over $\mH_d$ we get:
\begin{equation} \label{eq:factor}
 \exp_{\mH}(\theta_1 \bi_1 + \theta_2 \bi_2 + \dots + \theta_d \bi_d ) = \prod_j \exp_{\mH}(\theta_j \bi_j );
\end{equation}
in particular, the left hand side is a factorizable element. If we now define the $\mH_d$-valued array $F$ with entries
\[
F(k_1,\dots,k_d ; t_1, \dots, t_d) = \exp_{\mH}( 2\pi \sum_{j} \frac{k_j t_j}{T_j} \bi_j  ),
\]
then using identities (\ref{eq:multisum}) and (\ref{eq:factor}), we obtain the orthogonality relation

\begin{equation} \label{eq:orthogonality}
\sum_{t_1,\dots,t_d}  F(k_1,\dots,k_d ; t_1, \dots, t_d) F^\sharp(\ell_1,\dots,\ell_d ; t_1, \dots, t_d) = \left \{ 
\begin{array}{ll} 
\prod_j T_j  & k_1=\ell_1,\dots, k_d=\ell_d \\
0 & else 
\end{array}
\right .
\end{equation}
(Again the right side is a real element of $\mH_d$, as in the remark following (\ref{eq:exponsum})). We can now justify  correctness of the following definitions.

We define the hypercomplex Fourier transform $\hat{z} = \cF_d z$ of the $d$-dimensional hypercomplex array 
$z = (z_{k_1,\dots,k_d})$ via

\begin{eqnarray*}
\hat{z}_{t_1, \dots, t_d} &=& \frac{1}{\sqrt{\prod_{j=1}^d T_j}}  \sum_{t_1=0}^{T_1-1}   \dots \sum_{t_d=0}^{T_d-1} 
z_{k_1, \dots k_d}
\\ && \cdot  \exp(2 \pi \bi_1 \frac{k_1 t_1}{T_1})   \cdots  \exp(2 \pi \bi_d \frac{k_d t_d}{T_d}),
\end{eqnarray*}
where always the hypercomplex exponential $\exp_{\mH_d}$ is intended.

The inverse transformation $z = \cF_d^{-1} \hat{z}$  is defined by:
\begin{eqnarray*}
z_{k_1, \dots, k_d} &=& \frac{1}{\sqrt{\prod_{j=1}^d T_j}}  \sum_{k_1=0}^{T_1-1}   \dots \sum_{k_d=0}^{T_d-1} 
  \hat{z}_{t_1 \dots t_d}
\\ && \cdot  \exp(-2 \pi \bi_1 \frac{k_1 t_1}{T_1})   \cdots  \exp(-2 \pi \bi_d \frac{k_d t_d}{T_d}).
\end{eqnarray*}
The fact that $\cF_d^{-1}$ really is the inverse to $\cF$ follows from (\ref{eq:orthogonality}).

\section{Sampling and Recovery in Real Coordinates}
The central insight about multidimensional NMR
with States-Haberkorn-Ruben phase-sensitive detection (PSD) \cite{States82}
assumes an idealized model with no FID decay
and with spectral lines exactly at the specially chosen Fourier frequencies,
we can represent the spectrum to be recovered as a hypercomplex 
array $\bx = (x_{k_1,\dots,k_d})$. The insight is that
NMR physically implements the hypercomplex Fourier transformation $\cF_d \bx$.
Traditionally, 
recovery of the spectrum is effected by the inverse of this
transformation operator. 

According to the machinery developed
in the previous section, each FID $f_{t_1, \dots, t_{d-1}}$ 
observable at one single indel $(t_1, \dots, t_{d-1})$
is a hypercomplex-valued function of the direct time $t_d$: we write this as
\[
f_{t_1, \dots, t_{d-1}}(t_d) = (\cF_d \bx)_{t_1, \dots, t_{d}}, \qquad 0 \leq t_d < T_d.
\]
Underlying this hypercomplex-valued
time series are $2^{d}$ separate real-valued time series, say $z^j$,
available via the vector isomorphism $\phi$, by picking out specific 
coordinates of that vector:
\[
   z^j(t_d) = \phi(f_{t_1, \dots, t_{d-1}}(t_d))_j, \qquad 0 \leq t_d < T_d
\]
RPD and PCS propose that, at a single indel $(t_1, \dots, t_{d-1}$), 
we acquire, not the full hypercomplex FID
$f_{t_1, \dots, t_{d-1}}$ or equivalently the whole 
vector of time series
$\phi(f_{t_1, \dots, t_{d-1}})$
but instead a subset of the available real coordinate time series $(z^j : j \in J(t_1, \dots, t_{d-1}) )$,
where the subset-selector $J(t_1, \dots, t_{d-1})$ specifies some but typically not all
of the $2^{d}$ real coordinates.
For instance, in a two-dimensional experiment the FID belongs to $\mH_2$
and so has $4$ components. At each indel, RPD acquires
one out of the two complex reads, i.e. two of the four available real components. 


We now try to make the above undersampling operations very concrete. 
Let $N = \prod_j T_j$ be the total number of hypercomplex entries in the 
spectrum to be recovered. Concatenate together all the hypercomplex FID's 
$f_{t_1, \dots, t_{d-1}}(t_d)$ from all indels $(t_1, \dots, t_{d-1})$ with $0 \leq t_j < T_j$
 into a vector $\beff$ with $N$ hypercomplex coordinates.

Let $\cC$ denote the `coordinatization' operator which applies the vector isomorphism $\phi$
to every entry of $\beff$, producing in each case a $2^d$-vector and
concatenating all $N$ such vectors into a single $N \cdot 2^d$ vector $\bz = \cC(\beff) $ with real entries. 

Let $n = T_d \cdot \sum_{t_1,\dots,t_{d-1}} card(J(t_1,\dots,t_{d-1}))$
denote the total number of real-valued samples acquired by a given PCS subset-sampling schedule.
Concatenate together all the real-valued time series $(z^j : j \in J(t_1, \dots, t_{d-1}) )$
acquired from all the indels, producing a vector $\by$ with $n$ total real coordinates. Conceptually,
there is an $n$ by $N$ binary-valued selection matrix $\cS$ such that $\by = \cS\bz$.

We have the following pipeline from NMR spectrum $\bx$, to its 
FID's $\beff$, to its real-valued component time series $\bz$
and then to its acquired real-valued samples:
\[
   \stackrel[\mH-Spectrum]{}{\bx} \quad  \stackrel[{\mH}-Fourier]{\cF_d}{\mapsto} \quad \stackrel[{\mH}-FIDs]{}{\beff} \quad
   \stackrel[]{\cC}{\mapsto} \quad \stackrel[{\mR}-FIDs]{}{\bz} \quad \stackrel[]{\cS}{\mapsto} \quad \stackrel[\mR-Samples]{}{\by}  
\]
The end-to-end pipeline amounts to an {\it acquisition operator} $\cA$:
\[
    \cA = \cS \circ \cC \circ \cF_d .
\]
For computations, a real coordinate representation is ultimately necessary.
Let $\cC^{-1}$ denote the coordinate-to-hypercomplex mapping that takes the
concatenation of $N \cdot 2^d$ real numbers and delivers a hypercomplex vector with
$N$ entries. The pipeline $\cA \circ \cC^{-1}$ is conceptually a mapping between
a real vector space of dimension $N \cdot 2^d$ and another real vector space
of dimension $n$ and is in fact a
linear mapping, which can therefore be represented by an $n \times (N 2^d)$ matrix $A$.

\newcommand{\bs}{{\bf s}}
If we let $\bs = \cC(\bx)$ denote the real coordinatization of the
underlying hypercomplex spectrum,
the problem of recovering the spectrum from partial
component sampling is the same as the problem of solving the system of equations
\[
   \by = A \bs,
\]
for the unknown $\bs \in \bR^{N2^d}$, from measurements $\by \in \bR^{n}$.

Because $n < N 2^d$, 
$A$ has fewer rows than columns, and the system of equations is underdetermined,
justifying our use of the term `undersampling'. In such situations,
we can't hope to recover an accurate reconstruction of $\bs$ from $\by$ alone.
Fortunately, there is by now a considerable body of work, going back decades,
giving conditions whereby, if the underlying spectrum is not too `crowded', 
and if the columns of the matrix $A$ are sufficiently `incoherent', then 
approximate or even exact recovery is possible. The easiest to understand such
conditions assume that the columns of $A$ are normalized to unit
length and consider the quantity $\mu$ already introduced earlier
in (\ref{eq:cohdef}). They consider the minimum $\ell_1$
reconstruction rule
$$
(P_1)  \quad \min  \| \bs \|_{1} \quad \text{subject to} \quad A \bs= \by,
$$
where $A$ is an $n$-by-$N$ matrix,
$\by$ is an $n$-by-1 vector, and 
the minimization takes place over $\bs \in \bR^m$ and
$\|\bs\|_1 = \sum_{i} |s(i)|$. 
A simple result is this:
\begin{thm} \cite{DonohoElad02,Tropp04}
Suppose that $\by = A \bs_0$ where
$\bs_0$ has at most $k$ nonzeros.
Let $\bs_1$ denote some solution of $(P_1)$.
If  $k < (\mu^{-1}+1)/2$,
then $\bs_1=\bs_0$. 
\end{thm}
In short, low coherence allows exact recovery of sparse objects by convex optimization $(P_1)$.
Many interesting generalizations exist,
allowing much weaker sparsity conditions, other algorithms,
approximate sparsity, alternatives to coherence measurement,
and so on. 

In the context of NMR spectroscopy, the sparsity condition
is a quantitative way of saying that the spectrum is not too crowded
and the coherence condition is a way of saying that the sampling scheme
is sufficiently `diverse'.

Such existing results about the real-valued case, while suggestive to NMR practitioners,
fail to model a key ingredient of the problem: the hypercomplex
nature of the NMR experiment.  The results are not wrong, however
they only provide very weak information.
They work in a representation by real coordinates that expands 
by a factor $2^d$ the number of parameters to be estimated
and potentially also expands the number of nonzeros by a similar factor.

\section{Sparse Recovery in Hypercomplex Coordinates}

We now develop a sparse recovery result appropriate
for the hypercomplex setting, showing that
sufficiently sparse spectra can be exactly recovered
by an appropriate algorithm.

For a vector $\bx \in \mH_d^N$ with $N$ components $x(j)$ such that $\bx=(x(1),\dots ,x(N))$
the \emph{hypercomplex  one-norm} is\footnote{The hypercomplex
$1$-norm, if expressed in real coordinates,
applies the Euclidean norm to  $2^d$-dimensional vectors,
so it is a form of mixed $\ell_{2,1}$ norm.}  
$$
 \|X \|_{\mH,1} = \sum_{j = 1}^N | x(j) |_{\mH}. 
$$

Consider the optimization problem
$$
(P_{\mH,1})  \quad \min  \| 
\bx \|_{\mH,1} \quad \text{subject to} \quad \cA(\bx)= \by.
$$
This is an analog for the hypercomplex setting 
of  minimum $\ell^1$ optimization, which has
been successfully applied in numerous undersampling situations,
for example MR imaging \cite{CSMRI}. 

We say that a hypercomplex vector $\bx \in \mH_d^N$ is $k$-sparse
if there is a subset $I \subset \{1,\dots,N\}$ with $card(I) \leq k$ \-- the {\it support} \--
so that $x(j) = 0$, $j \not \in I$. For later use, we let $\bx^I$ denote the
subvector of $\bx$ consisting of those coordinates $i \in I$, and $\bx^{I^c}$ the subvector 
of the remaining coordinates.
With this notation, $\bx$ is $k$-sparse just in case for some $I$ with $card(I) \leq k$, $\bx^{I^c} = 0$.

To define and calculate the hypercomplex coherence, we employ properties of the acquisition matrix $A$
defined above, i.e. the $n$-by-$(N\cdot 2^d)$ real-valued matrix 
representing the end-to-end acquisition pipeline $\cA$. Let $\ba_j$ denote the $j$-th column of the matrix $A$.
The $i$-th entry of the hypercomplex vector $\bx$ has $2^d$ underlying real coordinates in
the real vector $\bx_0 = \cC(\bx)$, let's call this group of coordinates $K(i)$. For each pair $1 \leq i,j \leq N$,
let $G^{i,j}$ denote the $2^d$ by $2^d$ real matrix consisting of inner products
$(\ba_k)^T \cdot (\ba_{k'})$ with $k \in K(i)$, $k' \in K(j)$. For each such matrix $G = G^{i,j}$
let $\sigma_{max}(G)$ denote the largest singular value and $\sigma_{min}(G)$ the smallest singular value. 
Further, normalize the columns of matrix $A$ such that $\sigma_{min}(G^{i,i}) = 1 \text{\ \ for \ \ } 1 \leq i \leq N$, and let $G_n^{i,j}$ denote such a normalized version of $G^{i,j}$. 
We define the {\it hypercomplex coherence}
\[
  \mu_{\mH}(\cA) = \max_{i\neq j}  \sigma_{max}(G_n^{i,j})
  \]
In particular, if $\sigma_{min}(G^{i,i}) = 0$ for some $i$, we set $\mu_{\mH} = \infty$. 

Moreover, we define the (normalized) {\it hypercomplex point spread function (hPSF)} matrix 
\[
hPSF(i,j) = \sigma_{max}(G_n^{i,j})
 \]
For a general spike located at pixel $i$, $x_j^i = \delta_{i}(j)$, and generating measurement 
 $\by^i=\cA(\bx^i)$ and a matched-filter reconstruction $\cA^*(\by^i)$, a nonzero value of $hPSF(i,j)$ indicates a nonzero estimated coefficient at pixel $j\ne i$. Hence, one can use hypercomplex point spread function to identify the size and extent of undersampling-induced contaminations.


\begin{thm} 
 Let $\cA$ denote the end-to-end
sampling pipeline described in the previous section, with associated hypercomplex coherence $\mu_{\mH}(\cA)$.
Suppose we have measurements  $y  = \cA (\bx_0)$,
where $\bx_0 \in \mH_d^N$ is an $N$-dimensional 
$k$-sparse hypercomplex array.
 The solution to $(P_{\mH,1})$ is unique and equal to $\bx_0$ if 
$$
k < (1+ \mu_{\mH}^{-1})/2.
$$  
\end{thm}


\newcommand{\bv}{{\bf v}}


\begin{proof}
Let $A= \cA \circ \cC^{-1} \in R^{n\times (N2^d)}$ be the real sampling matrix associated with measurement operator $\cA$ as described in the previous section. Without loss of generality, we assume that the columns of the matrix $A$ are normalized such that 
$G^{i,i} = (\ba_k)^T \cdot (\ba_{k})$, \ \ $k \in K(i)$ has minimum singular value
equal to 1 for $1\leq i \leq N$. This is because $y = A x = ADD^{-1}x = A_n x'$ for a diagonal normalizing matrix $D$.

Suppose the solution to $P_{\mH,1}$ is not unique, and let $\bv = \bx_1 - \bx_0 \neq 0 $. 
As the solution of $(P_{\mH,1})$, $\bx_1$ must obey $\|\bx_1\|_{\mH,1} \leq  \|\bx_0\|_{\mH,1}$.
If $I$ denotes the support of $\bx_0$, then 
$\|\bx_1^I\|_{\mH,1} + \|\bx_1^{I^c}\|_{\mH,1}  \leq  \|\bx_0\|_{\mH,1}$. From the triangle inequality 
$ \|\bx_0^{I}\|_{\mH,1} -  \|\bx_1^{I}\|_{\mH,1} \leq \|\bx_1^I - \bx_0^I \|_{\mH,1} = \|\bv^I \|_{\mH,1}$ we get 
\begin{eqnarray*}
  \|\bv^{I^c}\|_{\mH,1} & \leq &  \|\bv^I\|_{\mH,1},
\end{eqnarray*}
and so
\beq \label{eq:boundonenorm}
 \|\bv^I\|_{\mH,1} \geq   \frac{\|\bv\|_{\mH,1}}{2}.
\eeq
In words, non-uniqueness requires that the hypercomplex 1-norm 
of the portion of $\bv$ on the support $I$ (i.e., $\bv^I$) must 
be at least half of the total hypercomplex 1-norm. 

\newcommand{\bu}{{\bf u}}
Because both $\bx_1$, and $\bx_0$ are 
assumed to be solutions to $\by = \cA \bx_0$, 
we have $A^T(A (\cC(\bx_1)-\cC(\bx_0)))= G \cC(\bv) =0$,
where $G$ denotes the $N 2^d$ square matrix made up of blocks $G^{i,j}$.
Hence
\begin{eqnarray*}
G^{i,i} \phi(v{(i)}) &=& - \sum_{j \neq i } G^{i,j} \phi(v{(j)}) \\
\| G^{i,i} \phi(v{(i)}) \|_2 &=& \| \sum_{j \neq i } G^{(i,j)} \phi(v{(j)}) \|_2 . \\
\end{eqnarray*}

Since for any real matrix $G$ and conformable real vector $\bu$, 
$\sigma_{min} \|\bu\|_2 \leq \| G \bu \|_2 \leq \sigma_{max} \|\bu\|_2 $,
and since $|z|_\mH = \|\phi(z) \|_2$,
\begin{eqnarray*}
\sigma_{min} (G^{i,i}) |v{(i)}|_\mH & \leq & \| G^{i,i} \phi(v{(i)}) \|_2 \\
&=& \| \sum_{j \neq i } G^{i,j} \phi(v{(j)}) \|_2 \\
&\leq &  \sum_{j \neq i } \sigma_{max} (G^{i,j})   \mid v{(j)}\mid_\mH .
\end{eqnarray*}
By normalization $\sigma_{min} (G^{i,i}) = 1 \  \ \forall i $, and so  
$|v^{(i)} |_\mH \leq   \mu  \cdot  \sum_{j \neq i } |v{(j)}|_\mH$ 
 and $| v^{(i)} |_\mH   \leq  \mu \left( \| v \|_{\mH,1} - | v^{(i)} |_\mH \right) $.
Summing across coordinates $i \in I$,
\begin{eqnarray*}
(1+\mu)  \sum_{i\in I }\mid v^{(i)} \mid_H &\leq&  {\mu \cdot \card(I)} \| v \|_{\mH,1},
\end{eqnarray*}
yielding
\beq \label{eq:boundoffsupport}
 \|\bv^I\|_{\mH,1} \leq \frac{\mu \cdot \card( I ) }{1+\mu}  \|\bv\|_{\mH,1}.
\eeq

Non-uniquness requires that (\ref{eq:boundonenorm}) and (\ref{eq:boundoffsupport}) both occur
simultaneously. This is not possible if
\begin{equation} 
\frac{\mu \cdot \card(I)}{1+\mu} < \frac{1}{2} \quad \mbox{i.e. } k < (1+ \mu^{-1})/2. \quad \qedhere \nonumber
\end{equation}
\end{proof}

\begin{figure*}[ht]
\centering
$
\begin{array}{lll}
\includegraphics[width=2in]{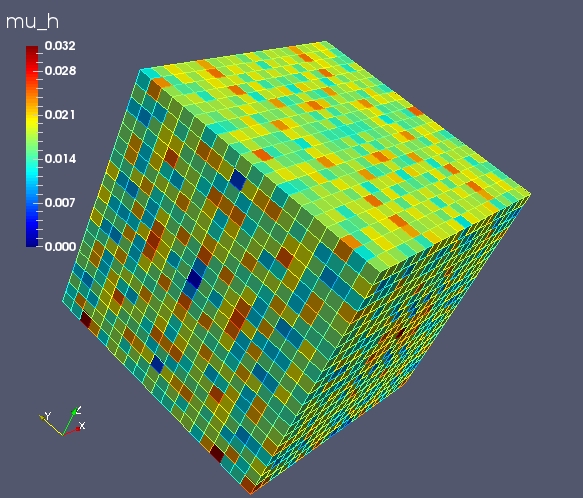}&
\includegraphics[width=2in]{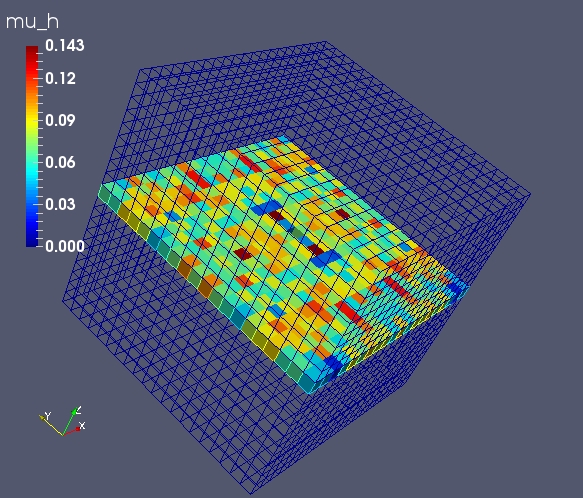}&
\includegraphics[width=2in]{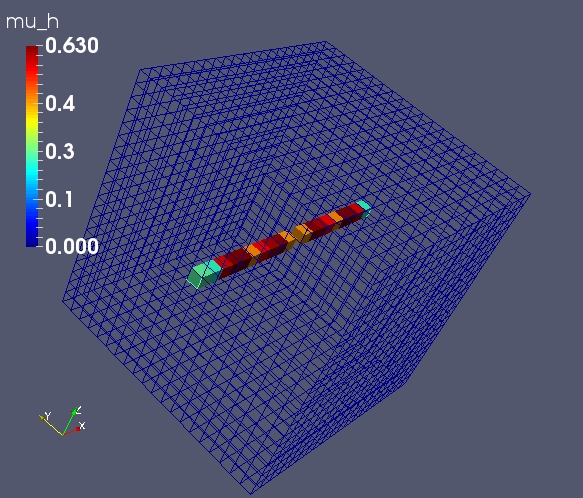} \\
\includegraphics[width=2in, height = 1in]{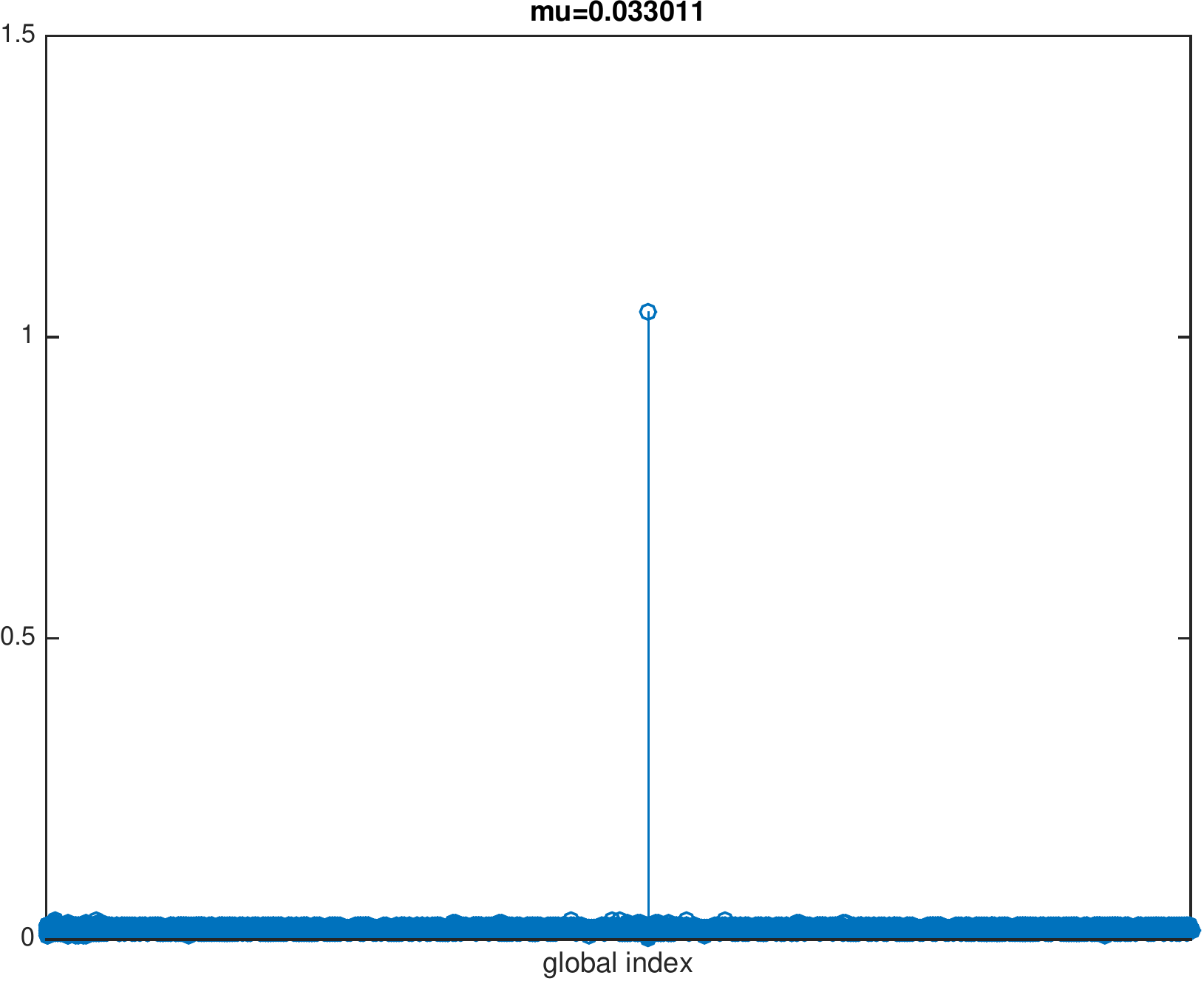}&
\includegraphics[width=2in, height = 1in]{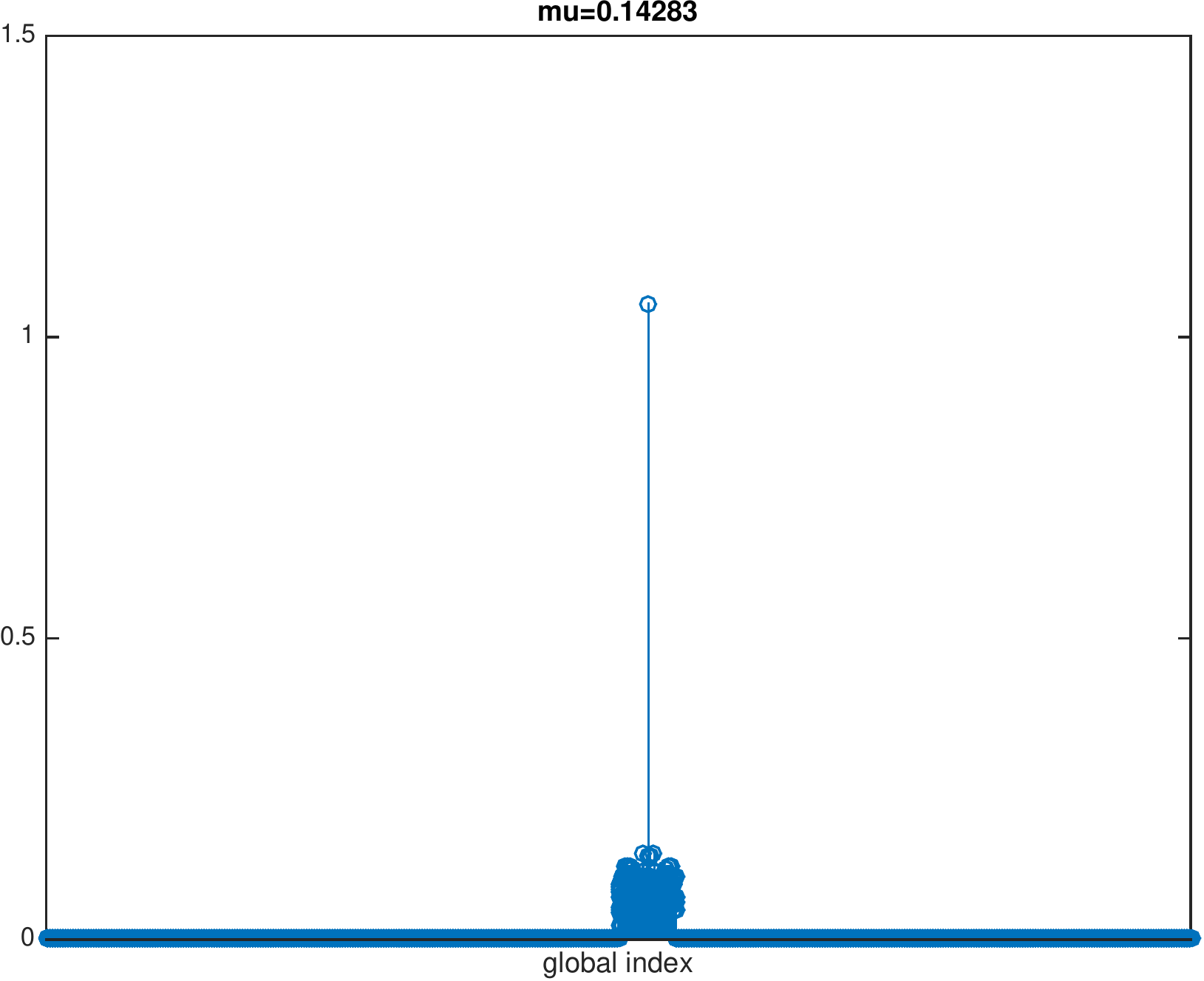}&
\includegraphics[width=2in, height = 1in]{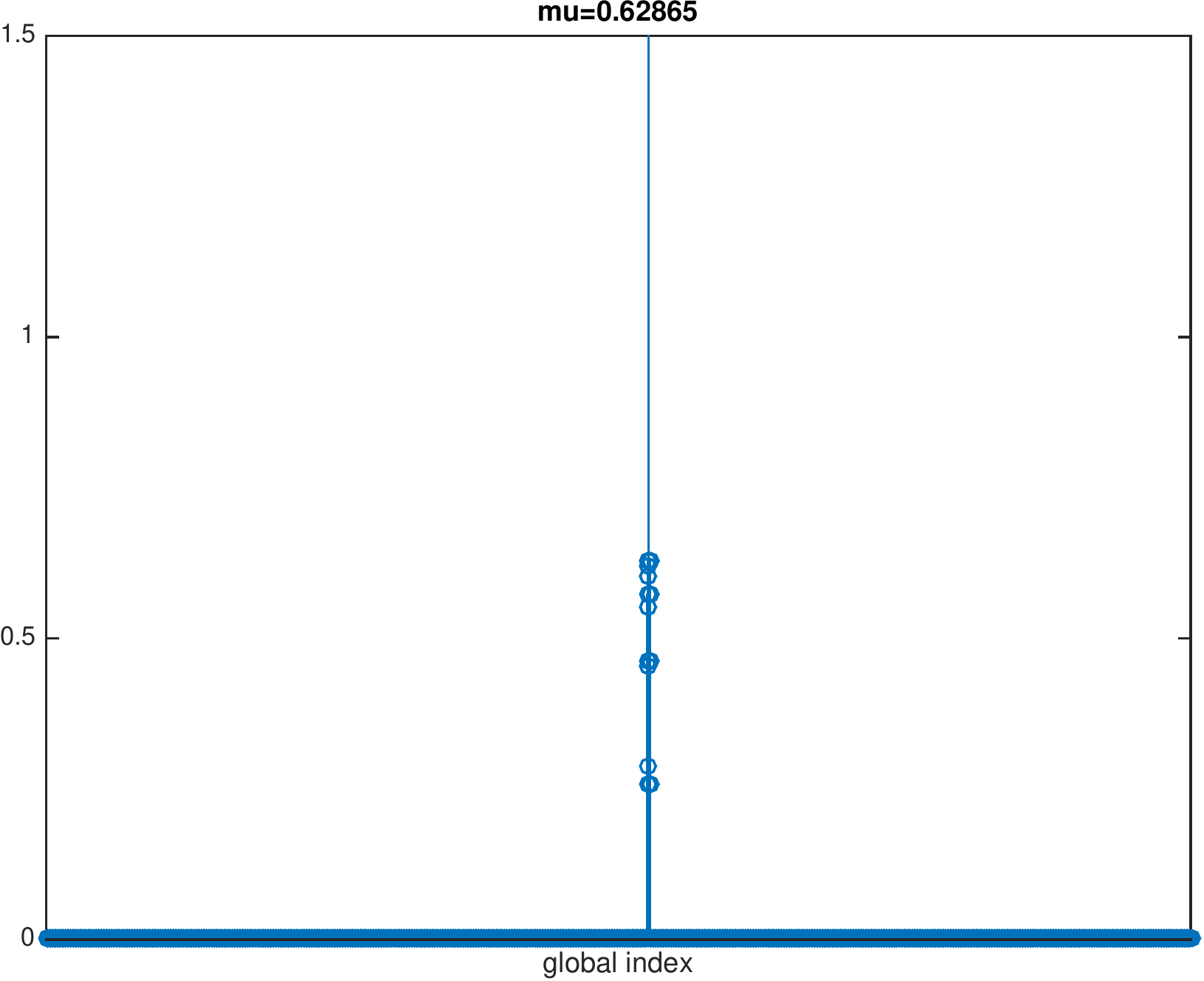} \\
\end{array}
$
\caption{hPSF corresponding to a spike at $i_0 = [0,0,0]$ and scheme S4 of partial component sampling for three different undersampling approaches 
A1 (pixels, left), A2 (indels, middle), and A3 (planes, right) in a three-dimensional problem of size $N=20\times20\times20$. 
The undersampling ratio $\delta=1/2$.
Each top panel has a different color scale for visibility purposes: 
(left) $[0,0.036]$, (middle) $[0,0.143]$, and (right) $[0, 0.711]$. 
The bottom panels showing the corresponding value of hPSF against an enumeration of the pixels. We see that more randomization of PCS schedule yields lower coherence.}
\label{fig-PCS}
\end{figure*}

\begin{figure*}[ht]
\centering
$
\begin{array}{lll}
\includegraphics[width=2in]{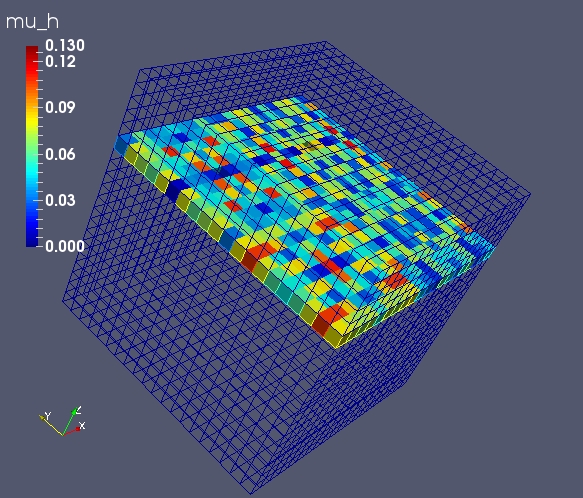}&
\includegraphics[width=2in]{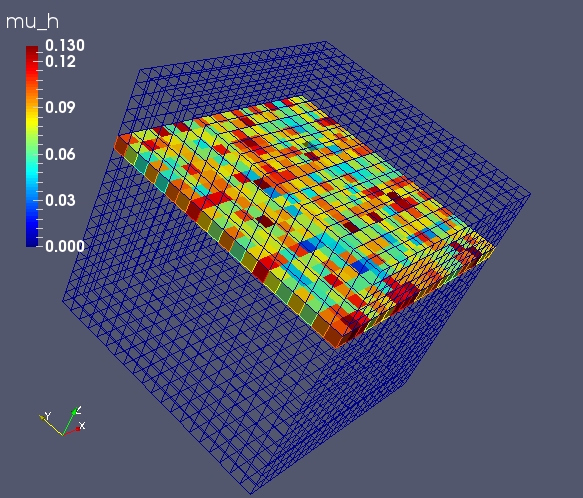}&
\includegraphics[width=2in]{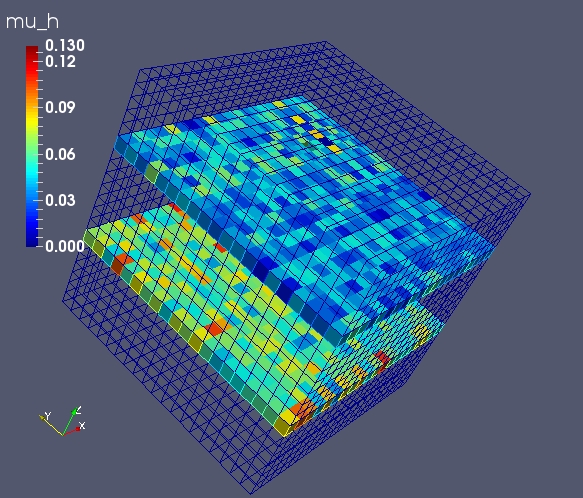} \\
\includegraphics[width=2in]{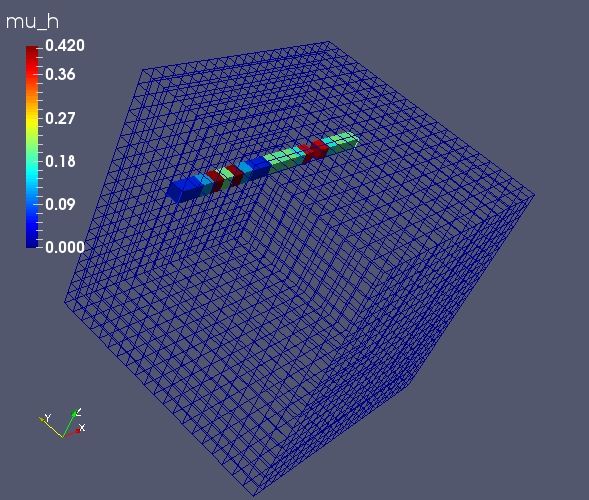}&
\includegraphics[width=2in]{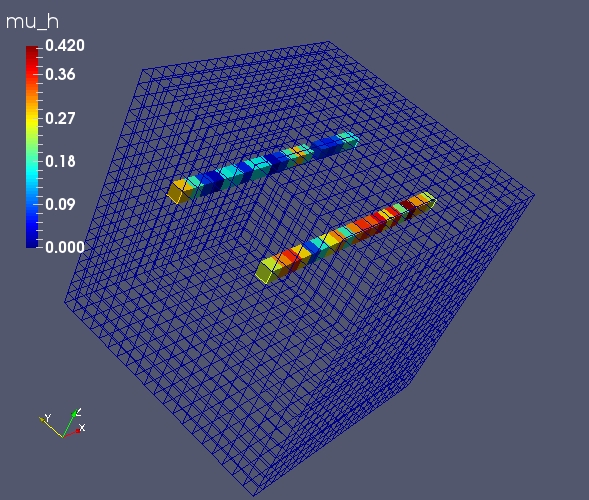}&
\includegraphics[width=2in]{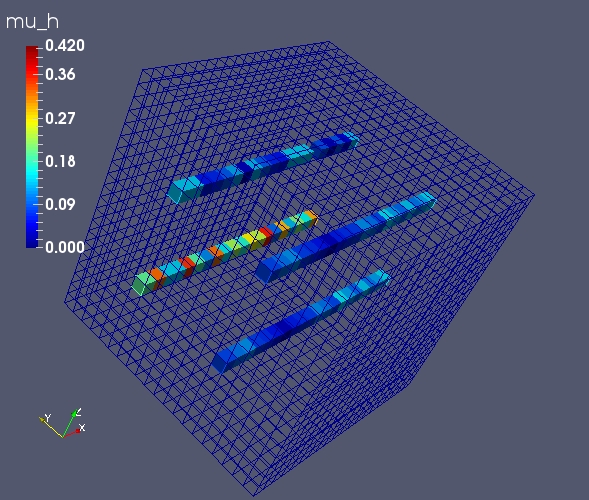} 
\end{array}
$
\caption{hPSF corresponding to a spike at $i_0 = [4,4,4]$ for NUS (left), FCPCS scheme S4 (middle), and FCPCS scheme S2 (right) in a three-dimensional problem of size $N=20\times20\times20$. Top and bottom panels correspond to approaches A2 (uniform sampling along indels) and A3 (uniform sampling along planes) respectively. We observe that, in general, NUS and PCS can produce different patterns of artifacts. Also, different schemes yield different contamination patterns. 
For NUS, the undersampling ratio $\delta=1/2$. For PCS, $\delta_i=1$, and  $\delta_c=1/2$.
Color scale for top panels $[0, 0.13]$ and for bottom panels $[0, 0.42]$.}
\label{fig-PCS-1684}
\end{figure*}

\begin{figure*}[h]
\centering
$
\includegraphics[width=3.5in]{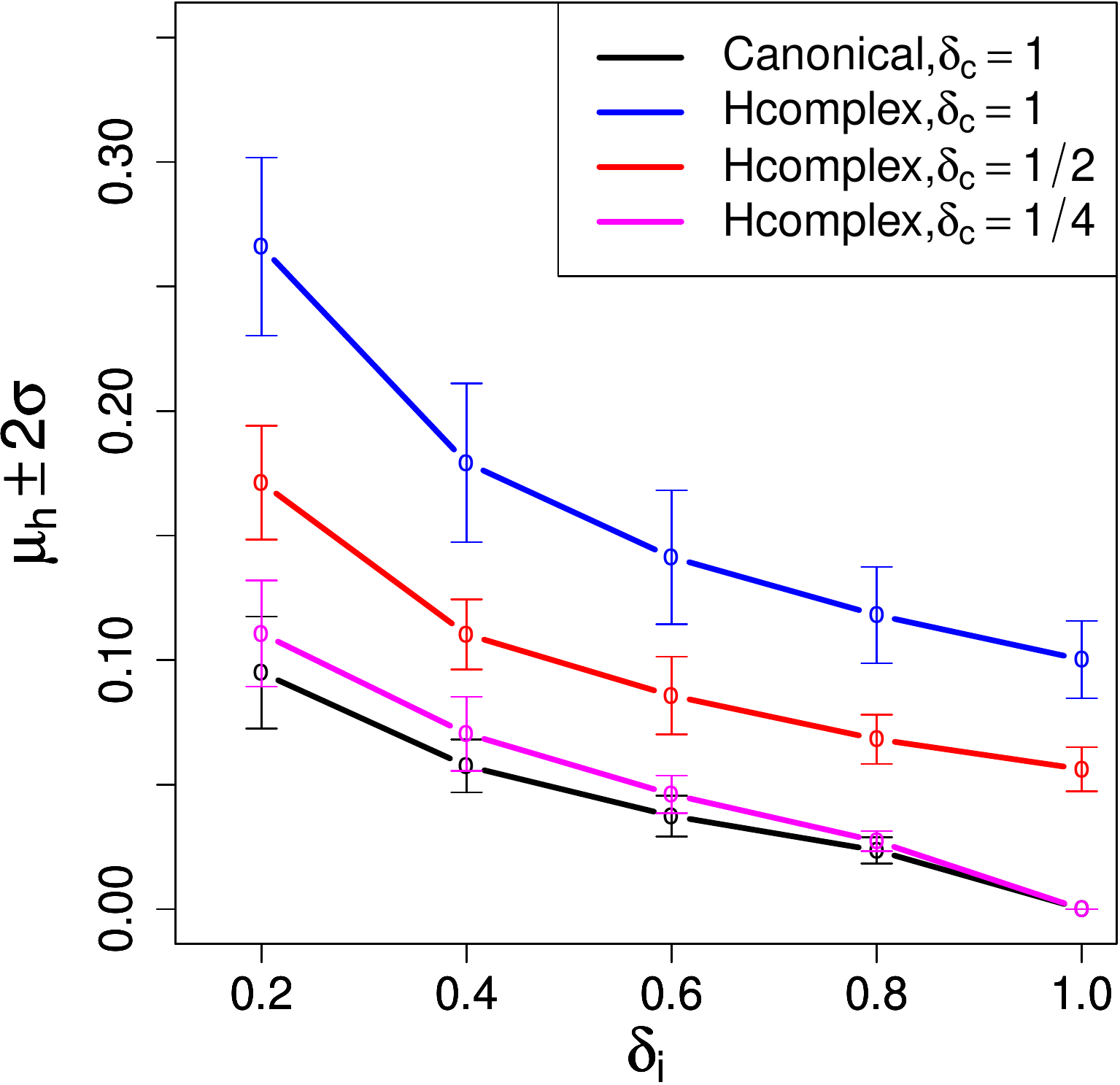}
$
\caption{Mean and standard deviation of coherence as a function of indel and component sampling coverage for a $50\times50$ indel grid and 10 Monte Carlo runs. The horizontal axis shows the indel undersampling ratio. Different colored lines correspond to the traditional measure of coherence for pure NUS (black), hypercomplex coherence for pure NUS (blue), hypercomplex coherence for a PCS approach when half of the hypercomplex components are sampled at selected indels (red) , and hypercomplex coherence for a PCS approach when a quarter of the hypercomplex components are sampled at selected indels (magenta).}
\label{mu-vs-sampling-coverage}
\end{figure*}

\begin{figure*}[h]
\centering
$
\includegraphics[width=3.5in]{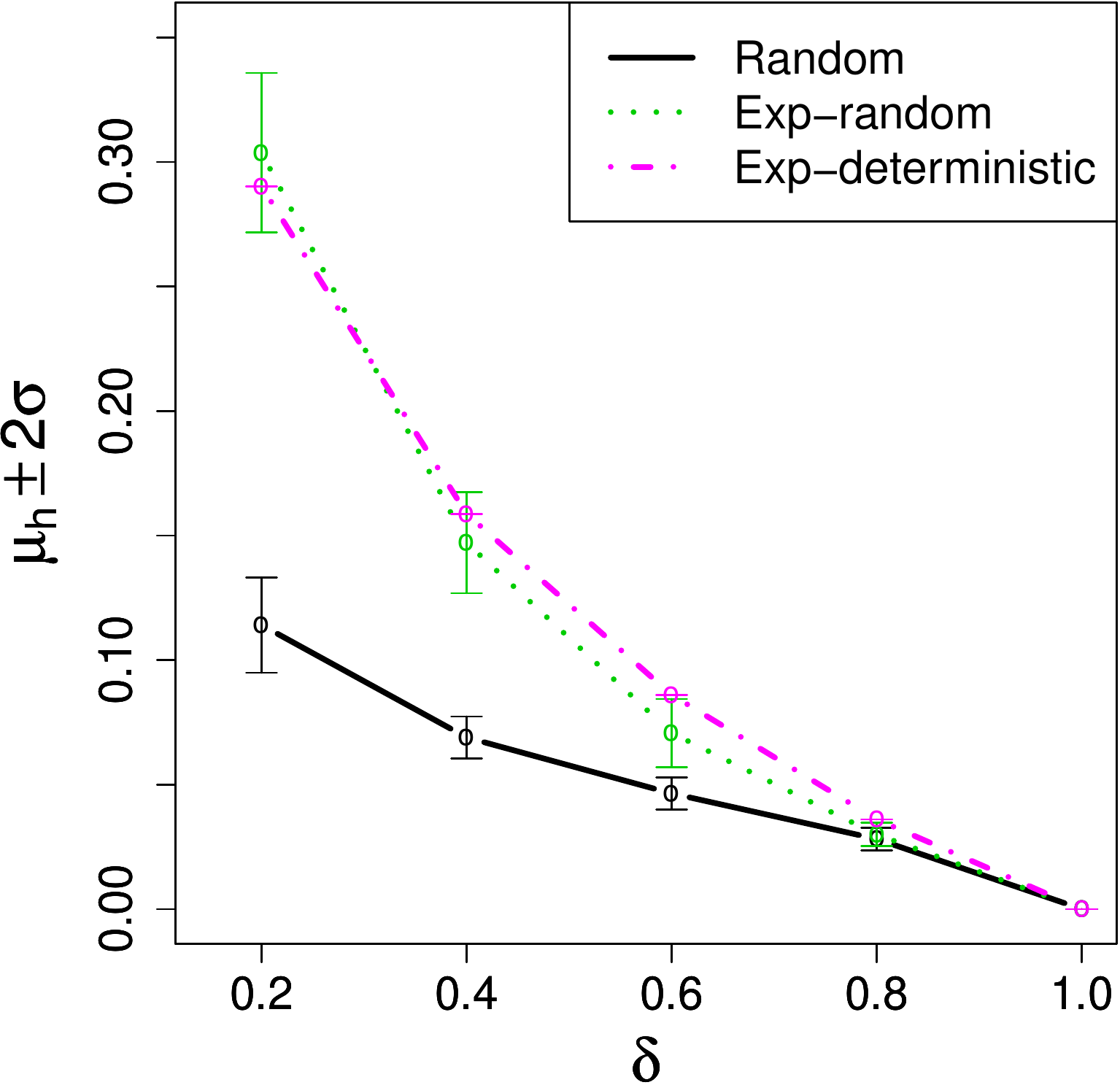}
$
\caption{Hypercomplex coherence for random (solid line), exponentially-biased random (dotted line), and deterministic exponential sampling (dash-dot line) for a $50\times50$ indel grid and 30 Monte Carlo runs. 
The horizontal axis shows the PCS undersampling ratio.}
\label{random-vs-exponential}
\end{figure*}

\begin{figure*}[ht]
\centering
$
\includegraphics[width=6in, height=2in]{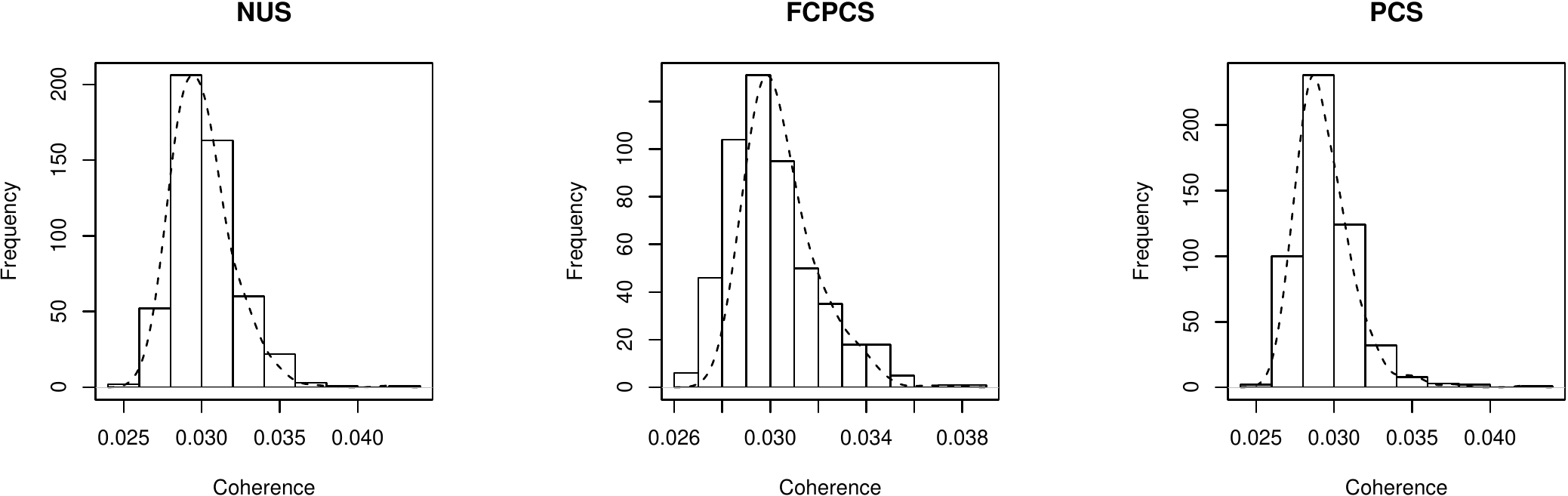}
$
\caption{Histogram of hypercomplex coherence for NUS (left), FCPCS (middle), and PCS (right) for undersampling ratio $\delta = 0.5$ and $100\times100$ indel grid. The dashed curve is a fitted smooth density function.} 
\label{coherence_dist}
\end{figure*}

\begin{figure*}[ht]
\centering
$
\includegraphics[width=6in, height=2in]{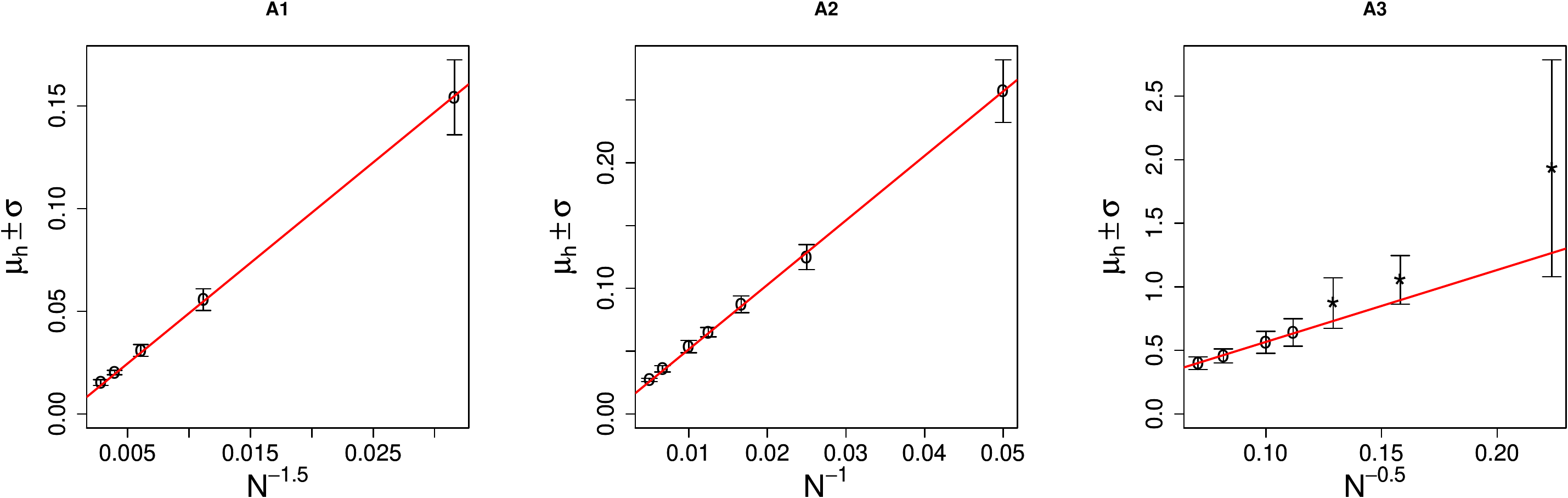}
$
\caption{Hypercomplex coherence for A1 (left), A2 (middle), and A3 (right) as a function of problem size in a three dimensional RPD experiment with undersampling ratio $\delta = 0.25$. The horizontal axis shows $N^{-(3-k)/2}$ where $k$ is the number of frozen dimensions for each approach. The red solid lines show the fitted lines in model (\ref{eq:fitted-model}). The fits are based on data for large enough problem sizes, i.e., $m=N^{d-k} > 60$ to eliminate the second-order effects associated with very small problems sizes. Data excluded from the fits are shown with $*$ symbols.}
\label{RPD_finiteN}
\end{figure*}

 \begin{figure*}[ht]
 \centering
 $
 \includegraphics[width=6in]{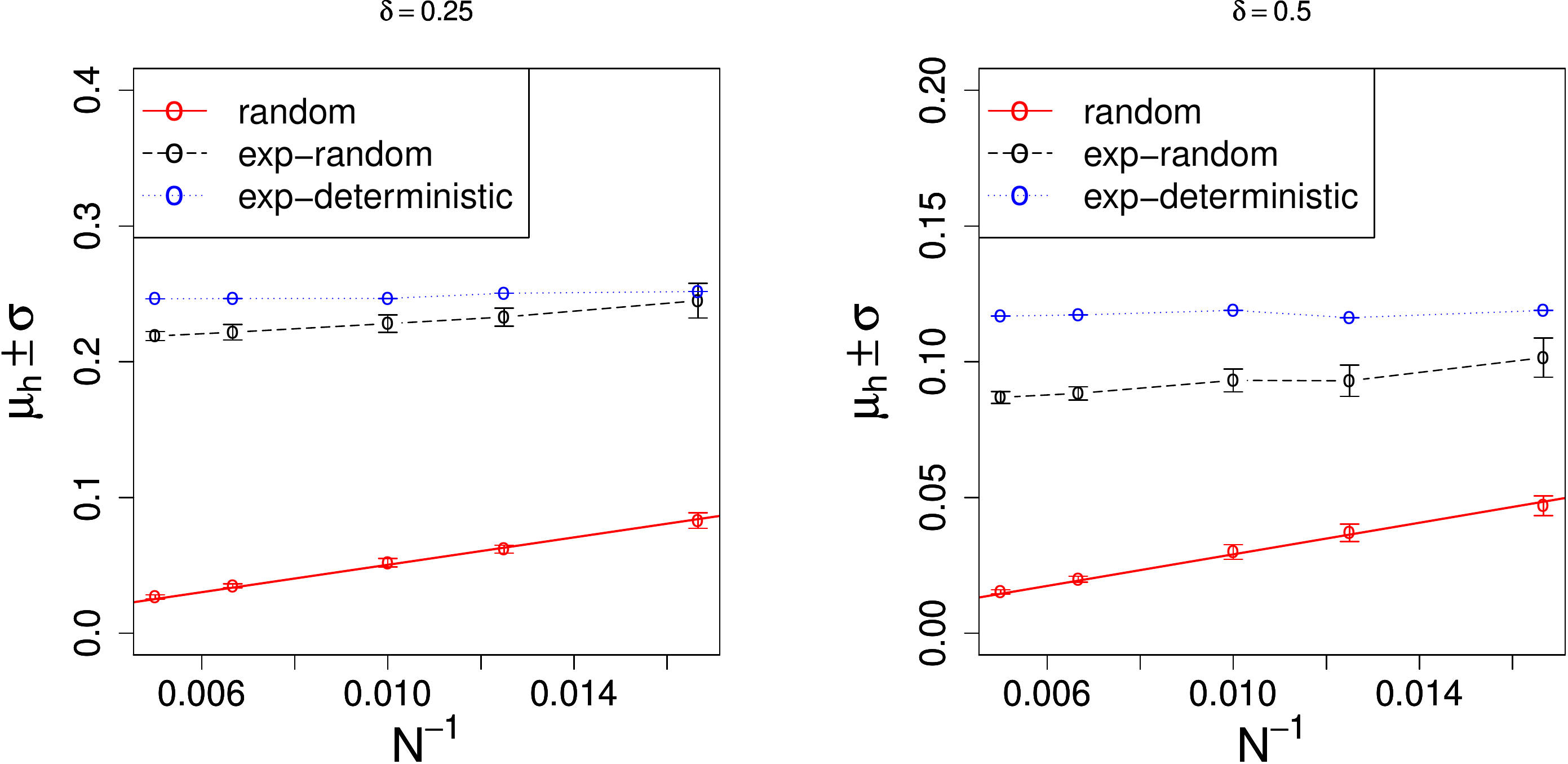}
 $
\caption{Hypercomplex coherence for random (red), exponentially-biased random (black), and deterministic exponential sampling (blue) as a function of problem size for undersampling ratio $\delta = 0.25$ and 30 Monte Carlo runs. The horizontal axis shows $1/\sqrt{m}$ where $m=N^2$ is the total number of indels. The solid red line shows the fitted line according to model (\ref{eq:intercept-free}).  } \label{PCS_finiteN}
\end{figure*}

\begin{figure*}[htbp]
\centering
$
\begin{array}{l}
 \includegraphics[width=5in]{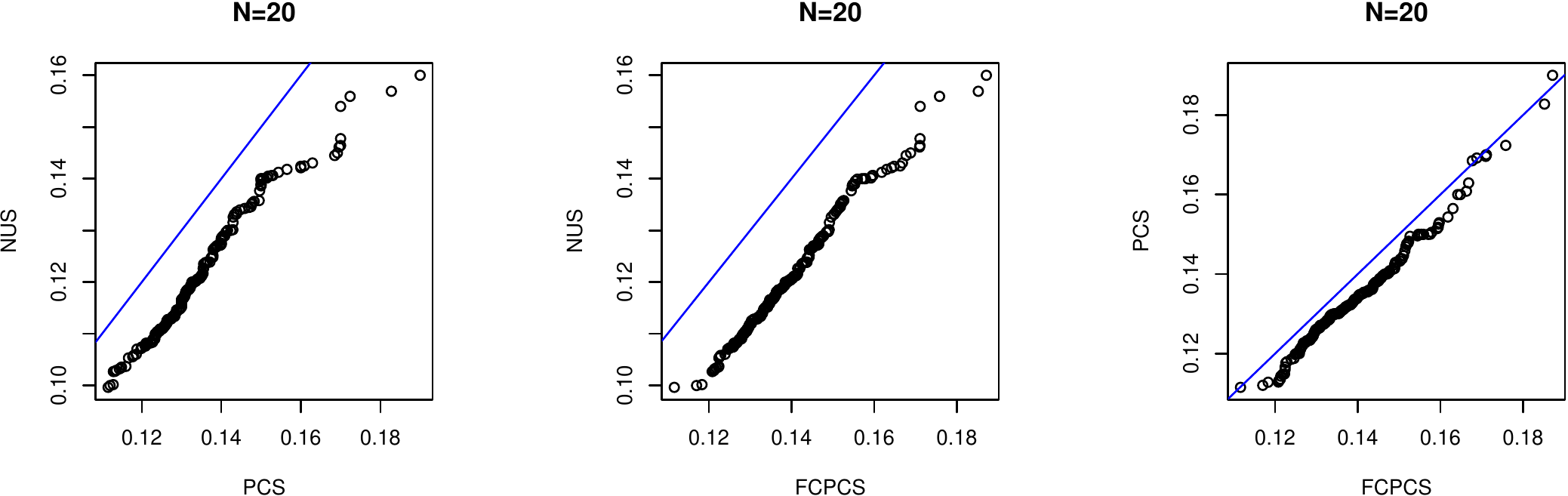}\\
\includegraphics[width=5in]{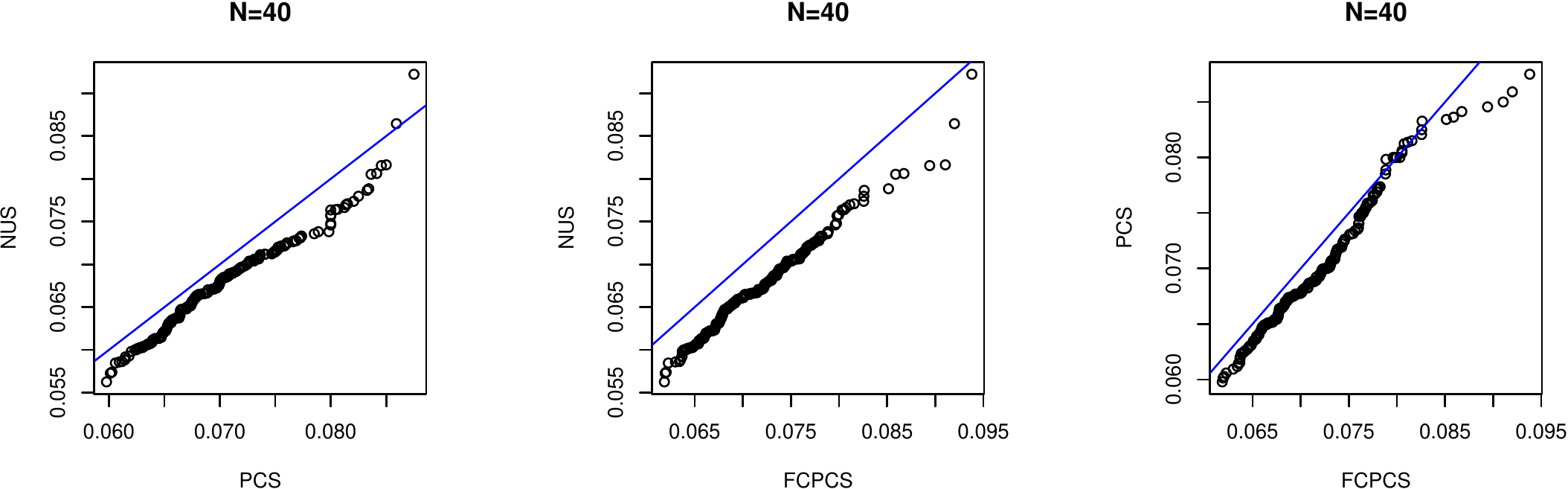}\\
\includegraphics[width=5in]{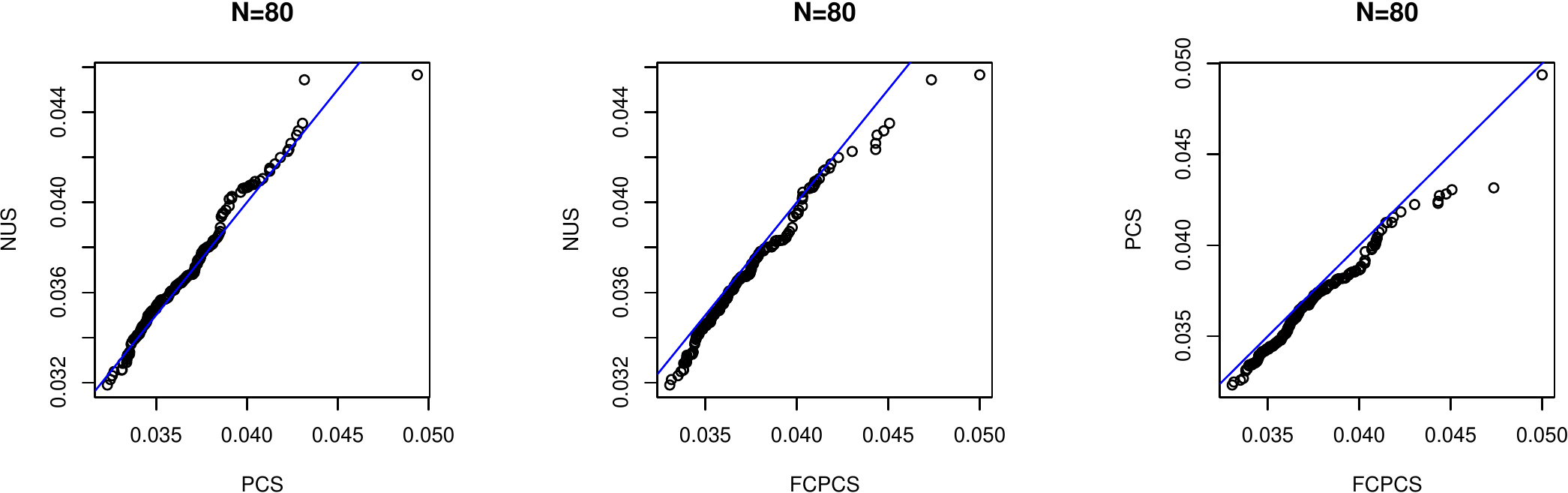}\\
\includegraphics[width=5in]{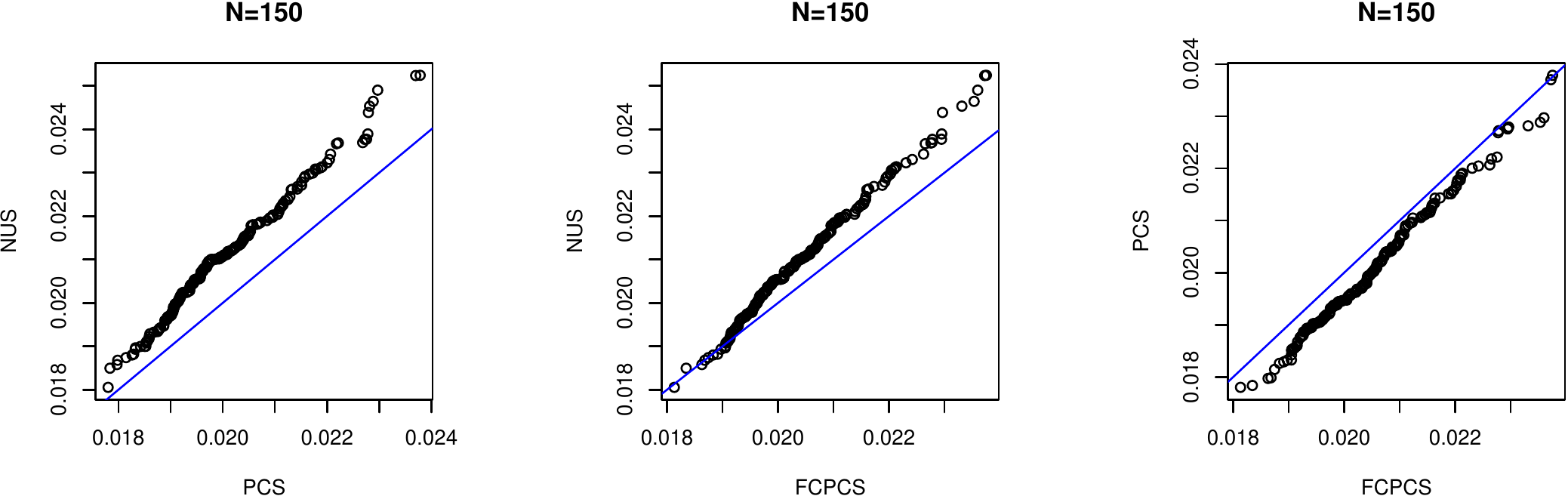}\\
\includegraphics[width=5in]{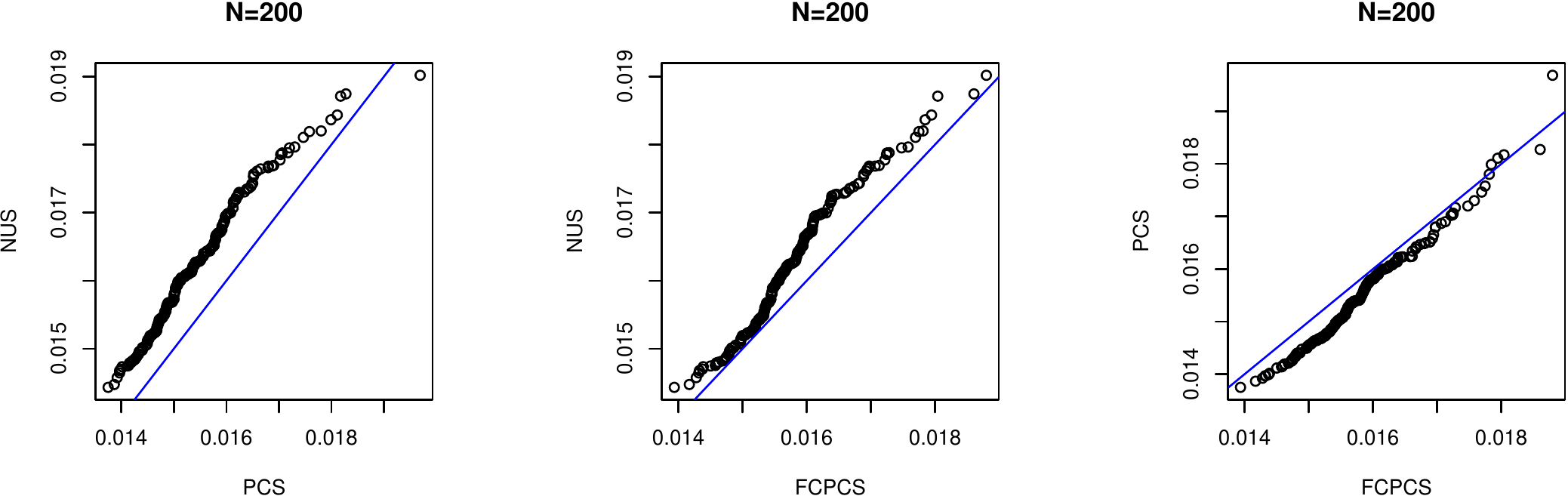} \\
\end{array}
$
\caption{Quantile-quantile plot of coherence distributions: NUS versus PCS (left), NUS versus FCPCS (middle), and PCS versus FCPCS (right) for approach A2 and scheme S4 on a 2D square indel grid of size $N=20,40,80,150,200$ (top to bottom). Here, $\delta=1/2$ and $nMonte=210$. In each panel, the blue line indicates the $y=x$ identity line. If the black dots all fall along the blue line, then the two distributions are equal. If they fall below the line, then the horizontally-plotted distribution is typically larger in value than the other one.}
\label{fig-qq-plot}
\end{figure*}

\section{Examples}

We now compute and interpret the hypercomplex coherence for three-dimensional NMR problems 
in which two indirect dimensions are undersampled. We consider these three types of experiments:
\bitem
\item {\itshape PCS with fixed component under-sampling ratio at all indels.} 
In this experiment, we fix the number of hypercomplex components measured at each indel. 
Here $0 \le \delta_i \le 1$ is the fraction of indels sampled. 
Similarly,  $0 \le \delta_{c} \le 1$ is the 
fraction of hypercomplex components sampled. 
We consider 8-dimensional 
hypercomplex numbers of the form 
$$
z = \left\{c_{0} + c_1 \bi_1 + c_{2} \bi_2 + c_{3}  \bi_{12} \right\}+ \bi_3   \left \{ c_{4} + c_{5} \bi_{1} + c_{6} \bi_{2} +  c_{7} \bi_{12} \right\}
$$

In our computations, we considered four \emph{schemes} of complex reads:
%
%

\begin{itemize}
\item $S1$ : $a \goto \left[c_0,c_1\right]$, $b \goto \left[c_2,c_3\right]$, $ c \goto \left[c_4,c_5\right]$, $d \goto\left[c_6,c_7\right]$
\item $S2$ : $a \goto \left[c_0,c_2\right]$, $b \goto \left[c_1,c_3\right]$, $ c \goto \left[c_4,c_7\right]$, $d \goto\left[c_5,c_6\right]$
\item $S3$ : $a \goto \left[c_0,c_3\right]$, $b \goto \left[c_1,c_5\right]$, $ c \goto \left[c_2,c_6\right]$, $d \goto\left[c_4,c_7\right]$
\item $S4$ : $a \goto \left[c_0,c_4\right]$, $b \goto \left[c_1,c_5\right]$, $ c \goto \left[c_2,c_6\right]$, $d \goto\left[c_3,c_7\right]$
\end{itemize}

For $\delta_c \in \{1/4,2/4,1\}$, and $s \in \left\{ S1, S2, S3, S4 \right\}$, we set the following sampling rules:
$$ 
z_{1/4}^{s} = \left\{
\begin{array}{ccc}
a[s]& \text{if} & \chi_{1/4} = 0\\
b[s] & \text{if} & \chi_{1/4} = 1\\
c[s] & \text{if} & \chi_{1/4} = 2\\
d[s] & \text{if} & \chi_{1/4} = 3\\
\end{array}
\right.
$$

$$ 
z_{2/4}^{s} = \left\{
\begin{array}{ccc}
\left( a[s], b[s] \right) & \text{if} & \chi_{2/4} = 0\\
\left( a[s], c[s] \right) & \text{if} & \chi_{2/4} = 1\\
\left( a[s], d[s] \right) & \text{if} & \chi_{2/4} = 2\\
\left( b[s], c[s] \right) & \text{if} & \chi_{2/4} = 3\\
\left( b[s], d[s] \right) & \text{if} &\chi_{2/4} = 4\\
\left( c[s], d[s] \right) & \text{if} & \chi_{2/4} = 5\\
\end{array}
\right.
$$
where $\chi_{\delta_c}$ denotes a random sample drawn from the integer set $\{0,1,\dots,(1/\delta_c-1)\}$ at fixed $\delta_c$.
According to these rules and through the following \emph{approaches}, we sample a subset of the hypercomplex components. 

\bitem
\item[+A1:]  $\chi_{\delta_c}$ is drawn once for each \emph{pixel} defined by  $(t_1,t_2,t_3)$.
\item[+A2:]  $\chi_{\delta_c}$ is drawn once for each \emph{indel} defined by $(t_1,t_2)$.
\item[+A3:]  $\chi_{\delta_c}$ is drawn once for each \emph{plane} defined by indirect index $t_1$. 
\eitem

Although mathematically well-defined, these sampling approaches 
and schemes are not equally realistic for NMR applications, and some are only
included in our analysis as instructive `sanity-checks'. 

\item {\itshape Non-uniform sampling.} 
In our NUS computtaions, all hypercomplex components are measured at the selected indels.  We consider the following selection schemes: 

\bitem
\item[+] {\bf Random sampling}: a subset of the indels is selected by uniform random sampling.  
\item[+] {\bf Exponentially-biased sampling}: indels at earlier times are more likely to be 
sampled than those at later times; the likelihood of sampling varies according to  an exponentially-decaying
probability schedule. Such sampling schemes are originally due to \cite{Barna87}. 
We  implemented both a deterministic and a random exponentially-biased sampling schedule. 
For  details, see the 
reproducible code distributed with this article (see section \ref{rr}).

\eitem 

\item {\itshape PCS with equal sampling coverage for all components.}
Here, we let $\delta$ represent the sampling coverage per hypercomplex component. 
For each hypercomplex component, we measure $\delta$ fraction of the indels through random or 
exponentially-biased sampling. 
\eitem

\subsection{Artifact patterns}
We now apply the hypercomplex point spread function to study the artifacts induced by undersampling  the hypercomplex Fourier transform. 
Figure \ref{fig-PCS} shows the hPSF for three different approaches of PCS at $\delta_c=1/2$, and $\delta_i=1$ due to a spike at $i_0 = [0,0,0]$. We observe a certain pattern of artifacts developing due to PCS, which can be understood using lemmas \ref{lemma} and \ref{lemma-PCS} in the Appendix. The plots are in agreement with the belief that allowing more randomization reduces the coherence \cite{CSMRI,hoch13}. In particular, restricting the randomization of PCS to the $t_1$ direction (i.e., A3) leads to a large nonzero point-spread along a line, while extending the randomization across two indirect dimensions $(t_1,t_2)$  (i.e., A2) leads to a somewhat smaller point spread throughout a plane. If we further allow randomization at every pixel (i.e., A1),  the point spread is much smaller, although nonzero throughout the
whole cube.

Figure \ref{fig-PCS-1684} shows the hPSF for a spike located at a nonzero frequency, namely $i_0 = [4,4,4]$. We observe that, in general, NUS and PCS develop different artifact patterns. However, for specific undersampling schedules, they may produce similar patterns (see top left and top middle panels). Also, we observe different patterns of artifacts for different schemes of PCS. For example, artifacts developed by scheme S4 are concentrated on positive frequency side (i.e., $i_z = 4$) whereas scheme S2 produces smaller size artifacts spread out on both sides of the frequency domain (i.e., $i_z = \pm 4$). Again, these patterns can be  understood using the lemmas presented in the Appendix.   

\subsection{Sampling coverage} 

Figure \ref{mu-vs-sampling-coverage} shows the coherence  
averaged over 10 Monte Carlo runs as a function of sampling coverage. 
The black line shows the traditional 
measure of coherence under a pure NUS scenario. 
The average hypercomplex coherence, similar to the traditional coherence, 
decreases as the sampling coverage increases. 
	
Figure \ref{random-vs-exponential} compares exponentially-biased sampling with  random sampling when two indirect dimensions are involved. As expected, we observe a lower coherence for random sampling schedules. 
	


\begin{table}[ht]
\centering
\caption{mean, standard error and Z-score of mean difference for NUS, FCPCS, and PCS coherence. Here, total undersampling $\delta = 0.25$, nMonte = 210 and $\delta_i = 1$ for FCPCS. The experiments are conducted for approach A2, and scheme S4 on a 2D square indel lattice of size N=40,60,80,100, and 200'. Standard errors are shown in parentheses. Numbers are rounded to the displayed precision.} 
\label{MethodEquiMean_delta2.5e-01}
\begin{tabular}{rlllll}
  \hline
 & 40 & 60 & 80 & 100 & 200 \\ 
  \hline
NUS & 0.117 (6.42e-04) & 0.083 (4.65e-04) & 0.064 (3.23e-04) & 0.052 (1.66e-04) & 0.028 (1.11e-04) \\ 
  PCS & 0.123 (7.62e-04) & 0.084 (4.47e-04) & 0.064 (3.23e-04) & 0.051 (1.45e-04) & 0.027 (1.01e-04) \\ 
  FCPCS & 0.127 (6.84e-04) & 0.085 (4.13e-04) & 0.065 (3.08e-04) & 0.053 (1.65e-04) & 0.027 (9.74e-05) \\ 
  Z(NUS,PCS) & -5.93 & -1.63 & 0.73 & 4.18 & 10.26 \\ 
  Z(NUS,FCPCS) & -11.17 & -4.18 & -1.41 & -3.92 & 6.79 \\ 
  Z(PCS,FCPCS) & -4.46 & -2.54 & -2.15 & -8.36 & -3.83 \\ 
   \hline
\end{tabular}
\end{table}

\begin{table}[ht]
\centering
\caption{mean, standard error and Z-score of mean difference for NUS, FCPCS, and PCS coherence. Here, total undersampling $\delta = 0.50$, nMonte = 210 and $\delta_i = 1$ for FCPCS. The experiments are conducted for approach A2, and scheme S4 on a 2D square indel lattice of size N=40,60,80,100, and 200'. Standard errors are shown in parentheses. Numbers are rounded to the displayed precision.} 
\label{MethodEquiMean_delta5e-01}
\begin{tabular}{rlllll}
  \hline
 & 40 & 60 & 80 & 100 & 200 \\ 
  \hline
NUS & 0.067 (3.85e-04) & 0.047 (2.23e-04) & 0.037 (1.79e-04) & 0.030 (9.37e-05) & 0.016 (6.63e-05) \\ 
  PCS & 0.070 (4.04e-04) & 0.048 (2.28e-04) & 0.036 (1.75e-04) & 0.030 (8.69e-05) & 0.015 (6.44e-05) \\ 
  FCPCS & 0.072 (4.04e-04) & 0.049 (2.75e-04) & 0.037 (1.84e-04) & 0.030 (8.25e-05) & 0.016 (5.62e-05) \\ 
  Z(NUS,PCS) & -4.71 & -1.47 & 0.73 & 5.62 & 8.33 \\ 
  Z(NUS,FCPCS) & -7.96 & -4.91 & -2 & 1.43 & 4.99 \\ 
  Z(PCS,FCPCS) & -3.17 & -3.55 & -2.75 & -4.51 & -3.94 \\ 
   \hline
\end{tabular}
\end{table}

\begin{table}[ht]
\centering
\caption{Coherence mean, standard error and Z-scores of mean difference for the four sampling schemes S1 to S4 in fixed-cardinality PCS. Here, total undersampling $\delta = 0.50$, nMonte = 200 and $\delta_i = 1$ for FCPCS. The experiments are conducted for approach A2 on a 2D square indel lattice of size N=20,40,60,80, and 100'. Standard errors are shown in parentheses.} 
\label{SchemeEquivalence}
\begin{tabular}{rlllll}
  \hline
 & 20 & 40 & 60 & 80 & 100 \\ 
  \hline
S1 & 0.136 (9.85e-04) & 0.070 (4.64e-04) & 0.047 (2.42e-04) & 0.036 (1.74e-04) & 0.029 (1.47e-04) \\ 
  S2 & 0.148 (9.94e-04) & 0.077 (3.86e-04) & 0.053 (2.44e-04) & 0.041 (1.96e-04) & 0.033 (1.41e-04) \\ 
  S3 & 0.150 (9.33e-04) & 0.078 (3.59e-04) & 0.053 (2.31e-04) & 0.041 (1.90e-04) & 0.033 (1.43e-04) \\ 
  S4 & 0.141 (1.00e-03) & 0.072 (4.29e-04) & 0.049 (2.45e-04) & 0.037 (1.74e-04) & 0.030 (1.32e-04) \\ 
  Z(S1,S4) & -3.504 & -2.779 & -4.795 & -4.645 & -4.727 \\ 
  Z(S2,S4) & 4.982 & 8.771 & 12.584 & 14.273 & 15.377 \\ 
  Z(S3,S4) & 6.728 & 10.14 & 12.897 & 14.093 & 14.114 \\ 
   \hline
\end{tabular}
\end{table}

\begin{table}[ht]
\centering
\caption{$P$-values of the estimated intercept $\hat{\beta}_0$ in model (\ref{eq:interceptModel}) for RPD.} 
\label{RPDfiniteN_pvals_delta2.5e-01}
\begin{tabular}{|c|lllllllll|}
  \hline
  & \multicolumn{9}{c|}{exponent $\gamma$} \\
 \hline approach &2 &1.75 &1.5 &1.25 &1 &0.75 &0.5 &0.33 &0.25 \\
 \hline
A1 & 0.021 & 0.020 & { \bf 0.050 } & 0.010 & 0.010 & 0.009 & 0.009 & 0.009 & 0.010 \\ 
  A2 & 0.002 & 0.002 & 0.001 & 0.001 & { \bf 0.221 } & 0.003 & 0.002 & 0.002 & 0.002 \\ 
  A3 & 0.002 & 0.002 & 0.001 & 0.001 & 0.000 & 0.000 & { \bf 0.292 } & 0.013 & 0.009 \\ 
   \hline
\end{tabular}
\end{table}

\begin{table}[ht]
\centering
\caption{$R^2$ of fits in model (\ref{eq:intercept-free}) for RPD. The exponent $\gamma = (d-k)/2$ gives the best fits for all the approaches.} 
\label{RPDfiniteN_Rqs_delta2.5e-01}
\begin{tabular}{|c|lllllllll|}
  \hline
  & \multicolumn{9}{c|}{exponent $\gamma$} \\
 \hline approach &2 &1.75 &1.5 &1.25 &1 &0.75 &0.5 &0.33 &0.25 \\
 \hline
A1 & 0.9767 & 0.9916 & { \bf 0.9997 } & 0.9941 & 0.9650 & 0.9028 & 0.8035 & 0.7201 & 0.6752 \\ 
  A2 & 0.8751 & 0.9122 & 0.9506 & 0.9838 & { \bf 0.9997 } & 0.9801 & 0.9069 & 0.8259 & 0.7775 \\ 
  A3 & 0.8598 & 0.8923 & 0.9242 & 0.9537 & 0.9781 & 0.9946 & { \bf 0.9999 } & 0.9959 & 0.9915 \\ 
   \hline
\end{tabular}
\end{table}

\begin{table}[ht]
\centering
\caption{$R^2$ of fits in model (\ref{eq:intercept-free}) for random PCS on 2D square indel grid and problem sizes $N=60,80,100,150,200$. The best fit occurs at $\gamma=1$.} 
\label{PCSfiniteN_Rqs}
\begin{tabular}{|r|lllllllll|}
  \hline
  & \multicolumn{9}{c|}{exponent $\gamma$} \\
 \hline $\delta$ &2 &1.75 &1.5 &1.25 &1 &0.75 &0.5 &0.33 &0.25 \\
 \hline
0.25 & 0.9193 & 0.9470 & 0.9714 & 0.9901 & { \bf 0.9995 } & 0.9955 & 0.9741 & 0.9486 & 0.9324 \\ 
  0.5 & 0.9153 & 0.9440 & 0.9694 & 0.9890 & { \bf 0.9992 } & 0.9959 & 0.9749 & 0.9497 & 0.9335 \\ 
   \hline
\end{tabular}
\end{table}

\begin{table}[ht]
\centering
\caption{$P$-values of the estimated intercept $\hat{\beta}_0$ in model (\ref{eq:interceptModel}) for random PCS on 2D square indel grid and problem sizes $N=60,80,100,150,200$.} 
\label{PCSfiniteN_pvals}
\begin{tabular}{|r|lllllllll|}
  \hline
  & \multicolumn{9}{c|}{exponent $\gamma$} \\
 \hline $\delta$ &2 &1.75 &1.5 &1.25 &1 &0.75 &0.5 &0.33 &0.25 \\
 \hline
0.25 & 0.004 & 0.003 & 0.003 & 0.003 & { \bf 0.049 } & 0.008 & 0.003 & 0.002 & 0.002 \\ 
  0.5 & 0.007 & 0.007 & 0.008 & 0.012 & { \bf 0.115 } & 0.001 & 0.001 & 0.001 & 0.001 \\ 
   \hline
\end{tabular}
\end{table}

\subsection{Comparison of NUS and PCS}

Figure \ref{coherence_dist} depicts the statistical distribution of hypercomplex coherence across many random realizations of undersampling. Here NUS, FCPCS, and PCS are compared, maintaining equal number of real degrees of freedom across schemes. The three methods exhibit visually similar \emph{skewed} distributions. 
It is very natural to test for statistical equivalence across these distributions. We begin by comparing the empirical \emph{average} coherence through the following hypothesis.

\hangindent=.5cm\textbf{Method Equivalence Hypothesis}. {\itshape Consider a three-dimensional NMR experiment with two indirect dimensions, each of length $N_i$, and a 
direct acquisition dimension of length $N_a$, leading to $N=2^3\cdot N_i^2N_a$ real coefficients. 
Suppose $n$ real coefficients are sampled 
 under each of three different random sampling methods $\cS_{NUS}$,  $\cS_{PCS}$ and $\cS_{FCPCS}$ and quadrature detection in the acquisition dimension (i.e., scheme $S4$).  The observed average coherence of PCS, FCPCS and NUS are the same to within sampling variation.}
 
We consider two forms of the Method Equivalence Hypothesis. A \emph{strong} form, which states the observed average coherences of NUS, PCS and FCPCS match for every $n$ and $N$, and a \emph{weak} form that says the differences of the observed average coherence decays to zero with increasing $N$.  
\begin{itemize}
 \item  \textit{Two-sample comparison}. We consider standard statistical procedures and work with the $Z$-score of mean difference between two samples: 
  	$$
	Z(\mu_0,\mu_1, M_0, M_1) = \frac{\mu_1 - \mu_2}{ \hat{SD}(\mu_0-\mu_1, M_0, M_1) }
	$$
Here $\mu_i$ denotes ``the observed average hypercomplex coherence for method $i$" and $\hat{SD}(\mu_1-\mu_2, M_1, M_2)$ is the appropriate standard error of comparing means for possibly unequal sample sizes $M_i$. We can now state the strong null hypothesis in terms of Z-scores:

\begin{itemize}
\item[+]\textbf{Strong Null Hypothesis.} The $Z$-scores of the mean differences are small for all $N$.
\end{itemize}

\item \textit{Asymptotic Equivalence.} It is not implausible to see significant differences in average coherence value at small $N$, and so the strict from of equivalence seems implausible a priori. Can we hope that the difference in average coherence becomes insignificant with increasing $N$? To investigate this possibility, we consider:

\begin{itemize}
\item[+]\textbf{Weak Null Hypothesis.} The $Z$-scores of the mean differences decrease with increasing $N$. 
\end{itemize}

\item \textit{Rejection of Method Equivalence Hypothesis.} 
Our results for testing the Method Equivalence Hypothesis are summarized in Tables \ref{MethodEquiMean_delta2.5e-01} and \ref{MethodEquiMean_delta5e-01} for undersmapling ratios $\delta=0.25$, and $\delta=0.5$. The $Z$-scores do not support the weak or strong form of the Method Equivalence Hypothesis although the difference of the average coherences between the methods seem to be inconspicuous.

\item \textit{Q-Q plots.} 
To further understand the relation between different sampling methods, we present quantile-quantile plots as shown in Figure \ref{fig-qq-plot}. Though the deviation from the identity line is small ($\sim <5\%$), we do see that PCS gives slightly smaller coherence for larger problem sizes ($N>80$) whereas NUS performs slightly better for smaller problem sizes ($N<80$). At  $N=80$, we observe no significant difference.  Moreover, we see that PCS always outperforms FCPCS by giving lower coherence. Though there is a subtle statistically observable difference between these methods, we must point out that from a practical point of view these differences seem unimportant (see Table~\ref{MethodEquiMean_delta5e-01} for a comparison).

 \end{itemize}
\subsection{Comparison of different schemes} 
We have introduced four different schemes $S1$ to $S4$ for partial component sampling. These schemes, though mathematically well-defined, are not equally realistic for the setup of NMR experiments in which quadrature detection is employed. Actually, only scheme $S4$ is realistic for quadrature detection because both $sine$ and $cosine$ components of the acquisition dimension (i.e., $t_d$) are sampled together. This is why we chose to work with this scheme in the Method Equivalence Hypothesis. 

It is mathematically instructive to test whether these schemes yield different coherence values if all other experimental parameters are kept intact. We therefore test the following hypothesis.

\hangindent=.5cm\textbf{Scheme Equivalence Hypothesis}. Consider a three-dimensional NMR experiment with two indirect dimension each of length $N_i$, and 
acquisition dimensions of length $N_a$ leading to $N=2^3\cdot N_i^2N_a$ real coefficients. Suppose $n$ coefficients are sampled 
 through four different schemes $S1$, $S2$, $S3$, and $S4$ and approach $A2$ (i.e., sampled component subset changes from indel to indel). 
The average hypercomplex coherence for all these four schemes are equal. 

Similar to the Method Equivalence Hypothesis, we consider the Z-score as a statistical measure of significance, and examine the strict and weak forms of 
Scheme Equivalence Hypothesis. Our results, shown in Table \ref{SchemeEquivalence}, reject both forms of the Scheme Equivalence hypothesis.

\subsection{Finite-N scaling} 

Consider an RPD experiment with $d$ dimensions each of length $N$. Further, suppose we freeze $k<d$ dimensions so that randomization of selected component subset is employed in only $d-k$ dimensions. For example in approach A3, $d=3$ and $k=2$. For large enough problem sizes, we propose the following finite-$N$ scaling model.

\beq 
\label{eq:fitted-model}
\mu(N) = \beta_1 N^{-(d-k)/2}.
\eeq

Here $\mu$ is the expected value of hypercomplex coherence. Figure \ref{RPD_finiteN} shows the data and fitted lines for the three approaches A1,A2, and A3.

\subsubsection{Justification of exponent.}
To justify the scaling law suggested in model (\ref{eq:fitted-model}), we consider a three-dimensional RPD experiment in which 2 out of 8 components are sampled according to three different approaches A1, A2, and A3. We examine the two following fitting models:

\beq 
\label{eq:interceptModel}
\beta(N) = \beta_0 +  \beta_1(\gamma) N^{-\gamma} + error
\eeq
and
\beq 
\label{eq:intercept-free}
\beta(N) = \beta_1(\gamma) N^{-\gamma} + error.
\eeq

The first model includes an intercept term allowing us to assess the possible significance of coefficient $\beta_0$. We examine the goodness of fit for three approaches of RPD as the exponent $\gamma$ varies. 

The $P$-values associated with coefficient $\beta_0$ of model (\ref{eq:interceptModel}) are reported in Table \ref{RPDfiniteN_pvals_delta2.5e-01}. The fitting exercise shows that 
$\gamma = (d-k)/2$ is the only exponent to make the case of $\beta_0 \neq 0$ very weak for all the approaches. 

Table \ref{RPDfiniteN_Rqs_delta2.5e-01} shows $R^2$ of fits in model (\ref{eq:intercept-free}). Evidently $\gamma = (d-k)/2$ gives the best fits for all the approaches.




\subsubsection{PCS finite-N scaling.}
So far we have shown that coherence of RPD as a function of problem size $N$ follows the scaling law given by model (\ref{eq:fitted-model}). Can we hope that the same model applies to random \emph{PCS with equal sampling coverage for all components}? 

We consider a three-dimensional NMR problem with two indirect dimensions and randomize the PCS schedule across the entire $N^2$ indels. Figure \ref{PCS_finiteN} shows the data of random PCS together with the fitted line in model (\ref{eq:intercept-free}). For comparison purposes, we show the data corresponding to exponentially-biased random and deterministic exponential sampling on the sample plot. Evidently, the \emph{random} case has lower coherence and decays faster with increasing problem size compared to the other two. 

Table \ref{PCSfiniteN_Rqs} shows $R^2$ of fits for the random case at two different undersampling ratios $\delta \in \{1/4,1/2\}$. Similar to the RPD case for approach A2, we see that the best fit occurs at $\gamma = 1$, verifying the adequacy of model \ref{eq:fitted-model} for general random PCS schedules.
Finally, Table \ref{PCSfiniteN_pvals} shows the $P$-values associated with coefficient $\beta_0$ of model (\ref{eq:interceptModel}) for  random PCS. Evidently, $\gamma = 1$ makes the evidence for $\beta_0 \neq 0$ the weakest.

\section{Discussion}
It has been shown that more randomization in a NUS schedule results in lower coherence \cite{hoch13}. We showed this to be equally true for hypercomplex-aware undersampling, in which a subset of all components are sampled. Our results show that coherence is the smallest when randomization of partial-component sampling is employed in as many indirect dimensions as possible.

Unlike \cite{Schuyler15}, the results of this article do not suggest a synergistic effect of PCS in reducing the sampling coherence compared to NUS; For a given fixed number of degrees of freedom for sampling, one observes almost the same coherence value no matter which way of expending the degrees of freedom is used. 
This seemingly contradictory observation is perhaps due to a different notion of coherence used in \cite{Schuyler15}. 


\section{Concluding remarks}
We presented a definition of coherence appropriate for the hypercomplex-valued setting of 
multi-dimensional NMR. Though the hypercomplex coherence is general,
in the sense that it can be applied to all undersampling classes, 
we would like to point out that this
measure is inherently different from the canonical notion of coherence for $d\ge 2$ even 
for the case of full-component random sampling in which one samples uniformly at random from
the collection of all indels, or even of all pixels.
This difference comes from the fact that in general 
$\sigma_{max}(\Phi(z)) \ne |z|_H$ for $z \in \mH_d$.\footnote{
The two quantities {\it are} equal over special subsets of $\mH_d$:
real, complex, and factorizable elements. The  inequality flows from aspects of the 
hypercomplex algebra $\mH_d$ to make it commutative. 
Other hypercomplex algebras such as quaternion and octonions behave differently. 
}


\section{Reproducible Research} \label{rr}
The code and data that generated the figures in this article may be found online
at \url{http://purl.stanford.edu/xn744fp3001}\cite{HcohereneData}.

\section*{Acknowledgments} 
We would like to thank Alan Stern for useful discussions on hypercomplex algebra. This work was supported by National Science Foundation Grant DMS0906812 (American Reinvestment and Recovery Act), and National Institutes of Health Grants R21GM102366 and P41GM111135.

\section*{References}
\bibliography{refs}

\section*{Appendix}

\begin{dfn}[uniformly-sampled dimension]
Consider a $d$-dimensional sampling set $\cT$, sampled from within a Cartesian grid of $T_1\times T_2, \dots, T_d$ equispaced points. We say that dimension $i \in \{1,2,\dots,d\}$ is uniformly sampled if for each $(d-1)$-tuple $t^{(i)}=(t_j:j\neq i)$ arising from a sample $t\in\cT$, every $d$-tuple $t^{(i)}(k)=(t_1,\dots,t_{i-1},k,t_{i+1},\dots,t_d), \ k=1,\dots,N_i$, occurs as a sample $t\in \cT$. 
\end{dfn}

\begin{lem}[NUS cross-correlation]\label{lemma}
Suppose that in the sampling set $\cT \in Z_{T_1}\times\dots\times Z_{T_d}$, the dimension indecis $u\in U$ are all sampled uniformly and sampling is full-component. Let $F_k(t)\in \mH_d$ denote the $(t,k)$ coefficient of the hypercomplex Fourier matrix for $t = (t_1, \dots, t_d)$, and $k = (k_1, \dots, k_d)$,
$$
F_k(t) = \frac{1}{\sqrt{\prod_{j} T_j}} \exp{(\sum_{j=1}^d 2\pi \bi_j k_j  t_j/T_j)},
$$
and $\rho_{k,\ell}\in \mH_d$ denote the cross-correlation between two distinct columns $k$, and $\ell$,

$$
\rho_{k,\ell} = \sum_{t \in \cT} F_k(t) F_{\ell}^\sharp (t). 
$$
Then, $\rho_{k,\ell} = 0$ unless $k_u = \ell_u  \text{\ \ for all \ \ } u \in U $

\end{lem}



\begin{proof}
The $(t,k)$ element of the hypercomplex Fourier matrix is

$$
F_k(t) = \frac{1}{\sqrt{\prod_{j} T_j}} \exp{(\sum_{j=1}^d 2\pi \bi_j k_j  t_j/T_j)}
$$
where $t = (t_1, \dots, t_d)$, and $k = (k_1, \dots, k_d)$.

Let $\cT$ denote the set of all tuples $t = (t_1,\dots,t_d)$
that get sampled.
Let $D = \{1,2,\dots,d\}$ represent the set of all dimensions. 
Further, let $U$ denote dimensions that are sampled uniformly and exhaustively
and $D\backslash U$ the rest that are sampled nonuniformly. For each index $j \in U$,
let $\cT_j = \{0,\dots, T_j-1 \}$ denote the full range of that index.
Let $\cT_{D \backslash U}$ denote the collection of all sampled indices 
in the non-uniformly-sampled variables
only. Then
\[
\cT =  \cT_{D \backslash U}  \times \prod_{j \in U} \cT_j.
\]

In the case of full-component sampling (i.e., NUS), the cross-correlation between two distinct columns $k$, and $\ell$ is given by, 

\begin{eqnarray*}
\rho_{k,\ell} &=& \sum_{t \in \cT} F_k(t) F_{\ell}^\sharp (t) 
\\ 
&=& \left( \frac{1}{\prod_{j \in D\backslash U} T_j}   \sum_{t \in \cT_{D\backslash U}} \exp{ ( \sum_{j\in D\backslash U} 2\pi \bi_j t_j  (k_j-\ell_j)/T_j ) } \right) \dots
\\
&&\left( \prod_{j\in U} \frac{1}{T_j} \sum_{t_j\in \cT_j} \exp{(2\pi \bi_j t_j  (k_j-\ell_j)/T_j)}\right)
\\
&=& \left( \frac{1}{\prod_{j \in D\backslash U} T_j}   \sum_{t \in \cT_{D\backslash U}} \exp{ ( \sum_{j\in D\backslash U} 2\pi \bi_j t_j  (k_j-\ell_j)/T_j ) } \right)  \cdot \prod_{j\in U} \delta(k_j-\ell_j)
\end{eqnarray*}
for uniformly sampled direction $u \in U$.

Here we used the exponential sum (\ref{eq:exponsum}) in each uniformly-sampled coordinate,
as well as the commutativity of the hypercomplex algebra, and $\delta(\cdot)$ denotes the Kronecker symbol.
We see that $\rho_{k,\ell} = 0 $ unless $k_u = \ell_u  \text{\ \ for all \ \ } u \in U $.
\end{proof}


So far we have shown that under \emph{full-component sampling}, if certain coordinates are sampled uniformly, then the cross-correlation between two distinct columns $k=(k_1, \dots, k_d)$ and $\ell=(\ell_1,\dots,\ell_d)$ of hypercomplex fourier matrix vanishes unless all their corresponding indecis of the uniformly-sampled coordinates match. When the cross-correlation does not vanish, one obtains a hypercomplex number $\rho_{k,\ell}$, which can be viewed as a $2^d$ by $2^d$ matrix $\Sigma_{k,\ell}=\Phi({\rho}_{k,\ell})$ where $\Phi$ is the matrix isomorphism defined for hypercomplex algebra in this article.

To generalize the results to partial-component sampling, we need to further examine the isomorphic matrix associated with correlation $\rho_{k,\ell}$ in detail. Before we proceed, it is helpful to define the quadrature acquisition in PCS.

\begin{dfn}[quadrature acquisition]\label{dfn:quadrature}
Let $F_k(t)\in \mH_d$ denote the $(t,k)$ coefficient of the hypercomplex Fourier matrix at $t = (t_1, \dots, t_d)$ and $k = (k_1, \dots, k_d)$,
$$
F_k(t) = \frac{1}{\sqrt{\prod_{j} T_j}} \exp{(\sum_{j=1}^d 2\pi \bi_j k_j  t_j/T_j)}.
$$
Each coefficient generates $2^d$ components given by,
$$
z_g =\frac{1}{\sqrt{\prod_{j=1}^d T_j}} \beta_1\times \dots 
\times \beta_d
$$
where 
$$\beta_j = \left\{ \begin{array}{cc}
\cos(2\pi k_j  t_j/T_j) & j\not \in g\\
\sin(2\pi k_j  t_j/T_j)& j \in g
\end{array} \right.$$ 
We say `quadrature acquisition' is employed in dimension $i$
if for each $(d-1)$-tuple $(\beta_j:j\neq i)$ arising from a sampling schedule of components $\cJ$, both components
$$\frac{1}{\sqrt{\prod_{j=1}^d T_j}} \left(\prod_{j\neq i} \beta_j\right) \cos(2\pi k_i  t_i/T_i), \quad \frac{1}{\sqrt{\prod_{j=1}^d T_j}} \left( \prod_{j\neq i} \beta_j\right) \sin(2\pi k_i  t_i/T_i)  $$ occur in $\cJ$.
\end{dfn}

Consider the end-to-end partial component sampling matrix denoted by $A$. Let us represent the entry of this matrix corresponding to frequency $k=(k_1,\dots, k_d)$ and sampled time $t=(t_1,\dots,t_d) \in \cT$ by 
$$a_{rc}^k(t), \quad r \in J(t), \ \  c=1,\dots, 2^d$$
where $J(t) \subset \{1,\dots,2^d\}$ is a set that contains the indecis of selected components at time $t$. 
It is easily verified that the entries of such sampling matrix takes the values
$$
a_{rc}^k(t) =\frac{\alpha^{rc}}{\sqrt{\prod_{j} T_j}} \beta_1^{rc}\times \dots \times \beta_d^{rc}
$$
for appropriate values of  
$\beta_j^{rc} \in \left\{ \cos(2\pi k_j  t_j/T_j), \sin(2\pi k_j  t_j/T_j\right\}$ and $\alpha^{rc} \in \{-1,1\}$. 
The $2^d\times 2^d$ cross-correlation matrix between two distinct columns $k$, and $\ell$ is  then given by
$$\Sigma_{u,v}^{k,\ell} = \sum_{t\in \cT} \sum_{r\in J(t)} a_{ru}^k(t) a_{rv}^{\ell}(t), \quad (u,v) \in \{1,\dots, 2^d\}^2.$$

To clarify, let us consider three different PCS scenarios in a two-dimensional NMR experiment in which one of the dimensions (direct dimension) is sampled uniformly. Namely,  
\begin{itemize}
\item (NUS):  $J(t) = \{1,2,3,4\},  \quad t_1 \in \cT_1, \  t_2 \in \{1,\dots,2^d\}$.
$$
a_{\{1:4,1:4\}}^k(t) = \frac{1}{\sqrt{T_1T_2}} \left(\begin{array}{cccc}
C_2C_1 &- C_2 S_1 & -S_2 C_1 & S_2S_1\\
C_2S_1 &  C_2 C_1 & -S_2 S_1 & -S_2C_1\\
S_2C_1 &- S_2 S_1 & C_2 C_1 & -C_2S_1\\
S_2S_1 &- S_2 C_1 & C_2 S_1 & C_2C_1\\
\end{array}\right) 
$$

\item (Quadrature PCS): $J(t) = \{1,3\},  \quad t_1 \in \cT_1, \  t_2 \in \{1,\dots,2^d\}$
   
$$
a_{\{1:2,1:4\}}^k(t) = \frac{1}{\sqrt{T_1T_2}} \left(\begin{array}{cccc}
C_2C_1 &- C_2 S_1 & -S_2 C_1 & S_2S_1\\
S_2C_1 &- S_2 S_1 & C_2 C_1 & -C_2S_1\\
\end{array}\right) 
$$

\item (Non-quadrature PCS):  $J(t) = \{1\}, \quad t_1 \in \cT_1, \  t_2 \in \{1,\dots,2^d\}$
$$
a_{\{1,1:4\}}^k(t) = \frac{1}{\sqrt{T_1T_2}} \left(\begin{array}{cccc}
C_2C_1 &- C_2 S_1 & -S_2 C_1 & S_2S_1\\
\end{array}\right) 
$$
\end{itemize}

In these examples $C_j=\cos(2\pi k_j  t_j/T_j)$ and $S_j=\sin(2\pi k_j  t_j/T_j)$ for $j=1,2$.

For (NUS) (i.e., full-component sampling), one can write
\begin{eqnarray*}
\Sigma_{1,2}^{k,\ell}  &=&  \sum_{t\in \cT}  \sum_{r=1}^4  a_{r1}^k(t) a_{r2}^{\ell}(t)  \\
&=& \sum_{t\in \cT}  (C_2C_1)^k (-C_2 S_1)^\ell + (C_2S_1)^k  (C_2 C_1)^\ell + (S_2C_1)^k (- S_2 S_1)^\ell + (S_2S_1)^k (- S_2 C_1)^\ell \\
&=& \sum_{t\in \cT}  - C_1^k S_1^\ell \left(C_2^k C_2^\ell + S_2^k S_2^\ell\right) +S_1^k C_1^\ell \left( C_2^k C_2^\ell + S_2^k S_2^\ell \right) \\
&=& \sum_{t\in \cT} \left( S_1^k C_1^\ell -C_1^k S_1^\ell \right)  \left( C_2^k C_2^\ell + S_2^k S_2^\ell \right) \\
&=& \sum_{t_1\in \cT_1} \left( S_1^k C_1^\ell -C_1^k S_1^\ell \right) \sum_{t_2\in \cT_2}  \left( C_2^k C_2^\ell + S_2^k S_2^\ell \right) \\
&=&\sum_{t_1\in \cT_1} \sin(2\pi (k_1-\ell_1)  t_1/T_1) \sum_{t_2\in \cT_2} \cos(2\pi (k_2-\ell_2)  t_2/T_2). 
\end{eqnarray*}
If one continues computing the other entries of cross-correlation matrix $\Sigma^{k,l}$, one sees that all non-trivial entries contain either $\sum_{t_j\in \cT_j} \sin\left(2\pi t_j  (k_j-\ell_j)/T_j \right)$ or $\sum_{t_j\in \cT_j} \cos\left(2\pi t_j  (k_j-\ell_j)/T_j \right) \text{\ \ for all \ } j \in D$. Therefore, $\Sigma^{k,\ell} = 0 $  unless $k_u = \ell_u  \text{\ \ for all uniformly sampled coordinate \ \ } u \in U$ and so we recover Lemma (\ref{lemma}) as expected.

For (quadrature PCS) case, we can write
\begin{eqnarray*}
\Sigma_{1,2}^{k,\ell}  &=&  \sum_{t\in \cT}  \sum_{r=1,3}  a_{r1}^k(t) a_{r2}^{\ell}(t)  \\
&=& \sum_{t\in \cT}  (C_2C_1)^k (-C_2 S_1)^\ell + (S_2C_1)^k (- S_2 S_1)^\ell \\
&=& \sum_{t\in \cT}  - C_1^k S_1^\ell \left(C_2^k C_2^\ell + S_2^k S_2^\ell\right) \\
&=&\sum_{t_1\in \cT_1}\cos\left(2\pi t_1 k_1 /T_1\right) \sin\left(2\pi t_1  \ell_1 /T_1\right) \sum_{t_2\in \cT_2} \cos(2\pi (k_2-\ell_2)  t_2/T_2). 
\end{eqnarray*}

Computing other non-trivial entries of the cross-correlation matrix in this case, one observes that they contain either $\sum_{t_j\in \cT_j} \sin\left(2\pi t_j  (k_j-\ell_j)/T_j \right)$ or $\sum_{t_j\in \cT_j} \cos\left(2\pi t_j  (k_j-\ell_j)/T_j \right) \text{\ \ for each quadrature acquisition dimensions \ } j \in Q$ (In the example above $Q={2}$). If, further, all the uniformly-sampled dimensions are acquired by quadrature detection (i.e., $U=Q$), we see that 
$\Sigma^{k,\ell} = 0 $  unless $k_j = \ell_j  \text{\ \ for all uniformly sampled coordinate \ \ } j \in U$ which can be considered as a generalization of Lemma (\ref{lemma}).

For the third scenario (non-quadrature PCS), we have 
\begin{eqnarray*}
\Sigma_{1,2}^{k,\ell}  &=&  \sum_{t\in \cT}  \sum_{r=1}  a_{r1}^k(t) a_{r2}^{\ell}(t)  \\
&=& \sum_{t\in \cT}  (C_2C_1)^k (-C_2 S_1)^\ell \\
&=& \sum_{t\in \cT}  - (C_1^k S_1^\ell) \left(C_2^k C_2^\ell \right)  \\
&=&\sum_{t_1\in \cT_1}\cos\left(2\pi t_1 k_1 /T_1\right) \sin\left(2\pi t_1  \ell_1 /T_1\right) \sum_{t_2\in \cT_2} \cos(2\pi k_2  t_2/T_2) \cos(2\pi \ell_2  t_2/T_2)
\end{eqnarray*}


Evidently, under non-quadrature PCS the factors do not necessarily contain $\sum_{t_j\in \cT_j} \sin\left(2\pi t_j  (k_j-\ell_j)/T_j \right)$ or   $\sum_{t_j\in \cT_j} \cos\left(2\pi t_j  (k_j-\ell_j)/T_j \right)$. Instead, the non-trivial entries of $\Sigma^{k,\ell}$ may contain the following
\begin{eqnarray}\label{trigIdentities}
\nonumber \sum_{t_j\in \cT_j}  \cos\left(2\pi t_j  k_j /T_j\right) \cos\left(2\pi t_j  \ell_j /T_j\right) & = & \sum_{t_j\in \cT_j}  \frac{1}{2}\left[ \cos\left(2\pi t_j  (k_j-\ell_j)/T_j \right) + \cos\left(2\pi t_j  (k_j+\ell_j)/T_j \right)\right]
\\
\nonumber \sum_{t_j\in \cT_j}  \sin\left(2\pi t_j  k_j /T_j\right) \sin\left(2\pi t_j  \ell_j /T_j\right) &=& \sum_{t_j\in \cT_j}  \frac{1}{2}\left[ \cos\left(2\pi t_j  (k_j-\ell_j)/T_j \right) - \cos\left(2\pi t_j  (k_j+\ell_j)/T_j \right)\right]
\\
\sum_{t_j\in \cT_j}  \sin\left(2\pi t_j  k_j /T_j\right) \cos\left(2\pi t_j  \ell_j /T_j\right) &=& \sum_{t_j\in \cT_j}  \frac{1}{2}\left[ \sin\left(2\pi t_j  (k_j+\ell_j)/T_j \right) + \sin\left(2\pi t_j  (k_j-\ell_j)/T_j \right)\right]
\end{eqnarray}

Therefore, under non-quadrature partial component sampling $\Sigma^{k,\ell} = 0 $ unless $k_u = \ell_u  \text{\ \ or \ \ }  k_u = T_u-\ell_u \text{\ \ for all \ \ } u \in U $ due to the exponential sum (\ref{eq:exponsum}) . We are now ready to establish the following lemma.

\begin{lem}[PCS cross-correlation]\label{lemma-PCS}
Suppose that in the sampling set $\cT \in Z_{T_1}\times\dots\times Z_{T_d}$, the dimension $u\in U$ are all sampled uniformly and sampling is partial-component. Let $D=\{1,\dots,2^d\}$ be the set of all dimensions and $Q \subset D$ represent all dimensions in which quadrature acquisition is employed. Further, let $A\in R^{n\times N2^d}$ denote the end-to-end partial component sampling matrix with entries corresponding to frequency $k=(k_1,\dots, k_d)$, and sampled time $t=(t_1,\dots,t_d) \in \cT$ given by

$$a_{rc}^k(t), \quad r \in J(t_{D\backslash U}), \ \  c=1,\dots, 2^d.$$
where $J(t_{D\backslash U}) \subset D$ is a set that contains the indecis of selected components corresponding to tuple $\left(t_j, \ j\not\in U\right)$. 
Further, let $\Sigma^{k,\ell} \in R^{2^d\times 2^d}$ denote the cross-correlation between two distinct columns $k$, and $\ell$
$$
\Sigma_{u,v}^{k,\ell} = \sum_{t\in \cT} \sum_{r\in J(t)} a_{ru}^k(t) a_{rv}^{\ell}(t), \quad (u,v) \in D\times D.
$$
Then, 
$\Sigma^{k,\ell} = 0$ unless 
$$
k_u = \left\{ \begin{array}{ll} \ell_u & u \in U \text{ \ and \ } u \in Q   \\
       \ell_u \text{ \ or \ } (T_u-\ell_u)  & u \in U \text{ \ and \ } u \not \in Q
\end{array}\right. .
$$
\end{lem}
\begin{proof}

The hypercomplex Fourier matrix is given by 
$$
F(t,k)  = \otimes_{j=1}^d  \left(\begin{array}{cc} \cos(2\pi k_jt_j/T_j) & - \sin(2\pi k_jt_j/T_j) \\ 
\sin(2\pi k_jt_j/T_j) & \cos(2\pi k_jt_j/T_j)
\end{array} \right).
$$
The end-to-end sampling matrix is a partially sampled hypercomplex Fourier matrix whose entries are given by

$$
a_{rc}^k(t) =\frac{\alpha^{rc}}{\sqrt{\prod_{j} T_j}} \beta_1^{rc}\times \dots \times \beta_d^{rc}, 
\quad r \in J(t_{D\backslash U}), \ \  c=1,\dots, 2^d ,
$$
for appropriate values of  $\alpha^{rc} \in \{-1,1\}$ and 
$\beta_j^{rc} \in \left\{ \cos(2\pi k_j  t_j/T_j), \sin(2\pi k_j  t_j/T_j\right\} \text{\ for all \ }j \in D$.
Then, the cross-correlation between two distinct columns $k$, and $\ell$ is 
\begin{eqnarray*}
|\Sigma_{u,v}^{k,\ell}| &=& \left | \sum_{t\in \cT} \sum_{r\in J(t_{D\backslash U})} a_{ru}^k(t) a_{rv}^{\ell}(t) \right |\\
&=& \left | \frac{1}{\prod_{j=1}^d T_j}  \sum_{t\in \cT} \sum_{r\in J(t_{D\backslash U})} \left( \alpha^{ru} \prod_{j=1}^d \beta_j^{ru} \right)_k  \left( \alpha^{rv} \prod_{j=1}^d \beta_j^{rv} \right)_\ell \right | \\
&=&  \left | \frac{1}{\prod_{j=1}^d T_j}  \sum_{r\in J(t_{D\backslash U})} (\alpha^{ru} \alpha^{rv}) \sum_{t\in \cT} \prod_{j=1}^d (\beta_j^{ru})_k (\beta_j^{rv})_\ell \right | \\
%
&=& \left| \sum_{r\in J(t_{D\backslash U})} (\alpha^{ru} \alpha^{rv})  \left\{ 
\left( \frac{1}{\prod_{j\in U \& j\in Q} T_j}  \sum_{t_j\in T_{j}}  \prod_{j\in U \& j\in Q} (\beta_j^{ru})_k (\beta_j^{rv})_\ell  \right) \right. \right. \dots \\  
&&\left( \frac{1}{\prod_{j\in U \& j\not \in Q} T_j}  \sum_{t_j\in T_{j}}  \prod_{j\in U \& j \not \in Q} (\beta_j^{ru})_k (\beta_j^{rv})_\ell  \right) \dots 
\\
&& \left. \left. \left( \frac{1}{\prod_{j\in D\backslash U} T_j}  \sum_{t_j\in \cT_{D\backslash U}}  \prod_{j\in D\backslash U} (\beta_j^{ru})_k (\beta_j^{rv})_\ell \right) \right\} \right|
\\
&=&(\frac{1}{2})^{\{\#j: j\in Q \& j\in U\}} \prod_{j\in U \& j\in Q} \delta(k_j-l_j) 
\prod_{j\in U \& j \not \in Q} \frac{1}{2} \left\{ \delta(k_j-l_j) \pm\delta(k_j+l_j) \right\} \dots
\\
&& \frac{1}{\prod_{j\in D\backslash U} T_j} \left| \sum_{r\in J(t_{D\backslash U})} (\alpha^{ru} \alpha^{rv})
\left\{\sum_{t_j\in \cT_{D\backslash U}} \prod_{j\in D\backslash U} (\beta_j^{ru})_k (\beta_j^{rv})_\ell  \right\}  \right|
\end{eqnarray*}
where we used trigonometric identities (\ref{trigIdentities}) and the exponential sum (\ref{eq:exponsum}) in the uniformly sampled coordinates together with commutativity of the hypercomplex algebra. In the `quadrature acquisition' dimensions, we expect to recover factors $\exp(2\pi k_j t_j/T_j), j \in Q$. We observe that $\Sigma_u,v^{k,\ell} = 0$ unless 
$$
k_u = \left\{ \begin{array}{ll} \ell_u & u \in U \text{ \ and \ } u \in Q   \\
       \ell_u \text{ \ or \ } (T_u-\ell_u)  & u \in U \text{ \ and \ } u \not \in Q
\end{array}\right.
$$
\end{proof}
Under exhaustive sampling, all coordinates are acquired by quadrature acquisition and  we recover Lemma \ref{lemma}.

\end{document}